\documentclass[twocolumn]{aastex631}

\usepackage[T1]{fontenc}
\usepackage[utf8]{inputenc}
\usepackage{float}
\newcommand{\bpic}[1]{$\beta$ Pic}
\newcommand{\apic}[1]{$\alpha$ Pic}

\usepackage{graphicx}
\usepackage{amsmath}
\usepackage{xcolor}

\shorttitle{}
\shortauthors{Avsar et al.}

\begin{document}

\title{Deep HST/STIS Coronagraphic Imaging of the $\beta$ Pictoris Debris Disk: On the Scattered Light Component of the Cat's Tail}

\author[0000-0001-7801-7425]{Arin M. Avsar}
\affiliation{Lunar and Planetary Laboratory, The University of Arizona, USA}

\author[0000-0002-4309-6343]{Kevin Wagner}
\affiliation{Steward Observatory, The University of Arizona, USA}

\author[0000-0003-3714-5855]{Dániel Apai}
\affiliation{Steward Observatory, The University of Arizona, USA}
\affiliation{Lunar and Planetary Laboratory, The University of Arizona, USA}
\affiliation{James C. Wyant College of Optical Sciences, The University of Arizona, USA}

\author[0000-0002-4388-6417]{Isabel Rebollido}
\affiliation{Centro de Astrobiolog\'ia (CAB, CSIC-INTA), Camino Bajo del Castillo s/n, 28692 Villanueva de la Ca\~nada, Madrid, Spain}

\author[0000-0003-4623-1165]{Antranik A. Sefilian}
\affiliation{Steward Observatory, The University of Arizona, USA}

\author[0000-0002-3191-8151]{Marshall D. Perrin}
\affiliation{Space Telescope Science Institute, 3700 San Martin Drive, Baltimore, MD 21218, USA}

\author{Christopher C. Stark}
\affiliation{NASA Goddard Space Flight Center, Exoplanets and Stellar Astrophysics Laboratory, Code 667, Greenbelt, MD 20771, USA}

\author[0000-0001-8009-8383]{Gabriel Weible}
\affiliation{Steward Observatory, The University of Arizona, USA}

\author[0000-0001-8612-3236]{András Gáspár}
\affiliation{Steward Observatory, The University of Arizona, USA}

\begin{abstract}

The $\beta$ Pictoris debris disk is a unique system where mid-infrared coronagraphic imaging with JWST/MIRI revealed a new spatially resolved substructure known as the Cat's Tail. We present deep HST/STIS coronagraphic imaging of $\beta$ Pic in scattered light, combining six epochs of observations between 2012 and 2025, in search of the scattered light component of the Cat's Tail. We detect the disk out to a projected separation of 500 au for the first time with HST/STIS, achieving SNR > 100 across much of the midplane between 50 and 200 au. We report the detection of the visible/near-infrared scattered light component of the Cat's Tail substructure. We find through injection-recovery analysis that the scattered light component of the Cat's Tail has a total flux of $0.1-0.4\%$ of the MIRI F1550C flux. Using the measured STIS-to-MIRI flux ratio, we model the Cat's Tail grain properties and find that the grains are highly porous and almost entirely composed of organic refractory material, which is in agreement with findings with JWST/MIRI. We use the organics content of Solar System dwarf planets and dust in the local ISM as a proxy to estimate the size of the  colliding progenitors that can produce enough organic refractory material seen in the Cat's Tail. We find that each colliding progenitor must have a mass of at least $1-3\times 10^{21}$ kg, comparable to Charon and Makemake in our Kuiper Belt. Additionally, we find a prominent warped morphology and surface brightness asymmetry in the outer dust halo of $\beta$ Pic. We compare the observed halo morphology and asymmetry to predicted vertical structures, which may arise from planet--disk interactions.

\end{abstract}

\keywords{}

\section{Introduction}

Multi-wavelength observations of debris disks probe grains of different sizes and temperatures, such that some dust populations are observable only at certain wavelengths \citep[e.g.,][]{Gaspar2023}. This wavelength dependence can itself be diagnostic, revealing previously unresolved substructures that trace dust populations distinct from the rest of the disk. Such structures can reveal not only compositional variation within the disk, but also dynamical activity, including in the form of recent collisions. The iconic $\beta$ Pictoris ($\beta$ Pic) debris disk is one such system in which compositionally distinct substructures have been discovered \citep[][]{Rebollido2024}.

$\beta$ Pic is an A6V main-sequence star located at a distance of 19.8 pc and has an age of $18.5^{+2.0}_{-2.4}$ Myr \citep[][]{Gaia2021,Miret-Roig2020}. It hosts a bright ($F_{disk}/F_{\star} \approx 2.5 \times 10^{-3}$; \cite{Lagrange2000}) and dynamically active debris disk that extends to separations of over 1000 au \citep[][]{Smith1984,Larwood&Kalas2001,Janson2021}. It is also one of the only known debris disk-hosting systems with confirmed super-Jupiter exoplanets: $\beta$ Pic b \citep[][]{Lagrange2009,Lagrange2010}, $\beta$ Pic c \citep[][]{Lagrange2019,Nowak2020}, and the newly discovered  $\beta$ Pic d \citep[][]{Gibbs2026,Sutlieff2026}. $\beta$ Pic b is known to influence the surrounding debris disk, with its presence originally predicted by the observed warp in the disk \citep[][]{Mouillet1997} and later confirmed with imaging \citep[][]{Lagrange2009,Lagrange2010}. The presence of a wide separation, massive planet with observational evidence of warping the surrounding dust and planetesimals likely enhances the collision rate of planetesimals in $\beta$ Pic \citep[e.g.,][]{Mustill&Wyatt2009}.

$\beta$ Pic is a dynamically active debris disk, with multiple lines of evidence for the continued dynamical upheaval of planetesimals resulting in major collision events occurring in the recent past. Mid-infrared imaging of $\beta$ Pic discovered a SW/NE brightness asymmetry due to an overdensity of dust at projected separations of between 50 to 80 au on the SW side of the disk \citep{Lagage1994,Pantin1997}. Follow-up mid-IR imaging from 8.7 $\mu$m to 24.6 $\mu$m by \cite{Telesco2005} found the spectral energy distribution (SED) of the SW clump to be consistent with dust grain properties that are hotter, smaller, and compositionally distinct from the rest of the disk. Due to the short blowout timescale of small grains, \cite{Telesco2005} argue that the aftermath of a recent massive collision can inject a distinct population of dust grains into the system. The dust clump on the SW side of $\beta$ Pic has since been observed from visible/near-infrared scattered light to sub-millimeter/millimeter observations \citep[e.g.,][]{Apai2015,Dent2014}. The collision hypothesis by \cite{Telesco2005} was further supported by the ALMA discovery of a CO gas clump co-located with the aforementioned dust clump in $\beta$ Pic \citep[][]{Dent2014}, as a massive planetesimal collision can sublimate CO ice on the colliding progenitors into the observed CO gas. Analysis of the orbital motion of the dust clump in the mid-IR by \cite{han2023} found that the SW dust clump is likely stationary, supporting the large planetesimal disruption hypothesis. \cite{skaf2023} conduct a similar orbital motion analysis of the clump and find that the clump has moved over time. The authors suggest an alternative explanation, with the dust clump being trapped by a gas vortex, but are unable to rule out the planetesimal disruption scenario.  

Mid-infrared observations of $\beta$ Pic with JWST has further supported the hypothesis that a planetesimal disruption occurred in the disk in the recent past \citep[][]{Chen2024,Rebollido2024}. Using coronagraphic observations centered at 15.5 $\mu$m and 23 $\mu$m, \cite{Rebollido2024} discovered an extended structure above the midplane of the SW side of disk, which they referred as the ``Cat's Tail", with the base of the tail being co-located with the dust and CO clumps previously discovered \citep[][]{Lagage1994,Pantin1997,Telesco2005,Dent2014}. The authors concluded that the Cat's Tail was created by an ongoing collisional cascade initiated by a collision between two approximately dwarf-planet-sized planetesimals in $\beta$ Pic $\sim$ 100-200 years ago.

In this study, we utilize new and archival observations of the $\beta$ Pic debris disk in search of the visible/near-infrared scattered light component of the Cat's Tail first discovered with the Mid-Infrared Imager (MIRI). We present deep coronagraphic imaging using the Space Telescope Imaging Spectrograph (STIS; \citealt{Kimble1998,Woodgate1998}) onboard HST. We combine archival datasets from multiple previous programs taken in 2012, 2021, and 2023 (published in \citealt{Apai2015,Avsar2024}) in addition to newly collected datasets taken in 2024 and 2025. We report the detection of the visible/near-infrared scattered light component of the Cat's Tail, and we use the flux limits of the Cat's Tail feature with HST/STIS to set compositional and geometric constraints on the properties of the dust within the Cat's Tail. In Section \ref{Observations-and-Reduction}, we outline the observations $\beta$ Pictoris with HST/STIS and associated data reduction steps. In Section \ref{Observational-Results} we highlight observational results of the combined new and archival observations of $\beta$ Pic. In Section \ref{Injection-Recovery}, we outline the injection/recovery technique used to model the flux of the Cat's Tail with STIS relative to MIRI. In Section \ref{Grain-Properties}, we outline the modeling of the grain properties of the dust in the Cat's Tail and report the results of the modeling. Finally, in Section \ref{discussion} we discuss the implications of detection of the Cat's Tail with STIS, the properties of the progenitors that created the Cat's Tail, and the origin of the asymmetry seen in the halo of $\beta$ Pic. 

\section{Observations and Data Reduction}\label{Observations-and-Reduction}

We present the reduction of six epochs of STIS observations of $\beta$ Pic. This includes three archival observations taken in 2012, 2021, and 2023 that have been presented in previous work by \cite{Apai2015} and \cite{Avsar2024}. We additionally include new STIS observations taken in 2024 and two epochs taken in 2025. Overall, we combine HST/STIS observations spanning 14 years, for a maximum pixel summed exposure time of $\sim 10^4$ seconds or $\sim$ 3 hours for $\beta$ Pic. Details on the observing configuration for each epoch of observation can be found in Appendix \ref{sec:STIS_obs}, including Tables \ref{tab:obs} and \ref{tab:obs-2}. All epochs were taken in the \texttt{50CORON} (coronagraphic) imaging mode and with identical occulter configurations. This mode has an unfiltered bandpass ranging from 0.2 to 1.1 $\mu$m \citep{STIS-DG}. For each epoch, $\alpha$ Pic was chosen to be the PSF reference star due to its similar color and apparent magnitude (V = 3.30, $\Delta$(B-V) w.r.t. \bpic{} = -0.01) and proximity on sky to $\beta$ Pic, allowing for minimal sun-spacecraft angle changes. Spacecraft rolls were performed during each orbit in order to image all sections of the disk that would have otherwise been blocked or obscured by the occulting wedges or diffraction spikes. The inner disk observations described in this section observed the disk from $\sim$10 to 200 au in projected separation, while the outer disk observations observed the disk from $\sim$60 to 500 au in projected separation.

\subsection{2012-2025 Inner Disk Observations}

The 2012 data were acquired as part of the program GO-12551 (PI: Apai). The observations were taken in three orbits, with the first and third orbits imaging $\beta$ Pic and the second orbit imaging the PSF reference star, $\alpha$ Pic. During each orbit, \texttt{WEDGEA0.6}, \texttt{WEDGEB0.6}, \texttt{WEDGEA1.0}, and \texttt{WEDGEB1.0} occulters were used to take short, medium, and long exposures. The short exposures used the \texttt{WEDGEA0.6} and \texttt{WEDGEB0.6} occulters to image the inner disk and the stellar PSF for both the target and reference star. The long exposures used \texttt{WEDGEA1.0} and \texttt{WEDGEB1.0} occulters to image the outer disk (past 2\farcs), while saturating the inner disk. 

The 2021, 2023, and 2024 epochs were acquired as a part of program GO-16174 (PI: Wagner). The data from both epochs were acquired in identical occulter and exposure time configurations to the 2012 epoch. The 2021-2024 epochs slightly varied from the 2012 epoch due to scheduling conflicts. Specifically, the 2023 epoch did not include the Visit 3 \texttt{WEDGEB0.6} exposures, and the 2021 and 2024 epochs reduced the total \texttt{WEDGEA0.6} exposures. 

The 2025 epoch of observations were acquired as a part of program GO-17741 (PI: Wagner). While these observations were intended to also be taken in an identical occulter configuration as the 2012 epoch, the Reduced Gyro Mode (RGM) of HST required changes to the observing plan in order to be executable. Specifically, the disk was observed only using the \texttt{WEDGEA1.0} occulter, removing exposures using the \texttt{WEDGEA0.6}, \texttt{WEDGEB0.6}, and \texttt{WEDGEB1.0} occulters. This still allowed for comparable total exposure times in the same regions of the disk as the previous observations.

\subsection{2023-2025 Outer Disk Observations}

The observations of the outer regions of the $\beta$ Pic disk ($>$ 200 au) were taken as a part of program GO-17456 (PI: Wagner). The observations were taken using a total of 17 orbits. These observations are different than the inner disk observations as they aim to detect the low surface brightness outer region of the disk. To achieve this, $\beta$ Pic was observed with long exposures and wider \texttt{WEDGEA1.8} and \texttt{WEDGEB1.8}. Most importantly, the absolute orientation of HST was set such that the disk's maximum extent is visible along the line joining the occulting wedge position and the corner of the detector. This method saturated the inner disk until approximately 60 au in projected separation but allowed for the high signal-to-noise ratio (SNR) detection of the outer disk out to 500 au for the first time with STIS.

\subsection{HST/STIS Coronagraphic Data Reduction}

The reduction of the raw coronagraphic images of the $\beta$ Pic disk closely follows the reduction technique outlined in \cite{Apai2015} and \cite{Avsar2024}. 

We start by downloading the cosmic ray rejected and flat fielded files, {\tt\string crj}, from the Mikulski Archive for Space Telescopes (MAST) Portal. After background subtraction and unit converion to mJy arcsec$^{-2}$ has been conducted on each image, we precisely locate the position of the star behind the occulting wedge using {\tt\string centerRadon} for both target and PSF reference images \citep{Ren2017}. Each image is then shifted to a common center using  {\tt\string scipy.ndimage.interpolation}. 

We then conduct PSF subtraction of each target image using the PSF reference star images of $\alpha$ Pic. To best subtract the stellar PSF and introduce minimal systematics, we conduct a three-dimensional grid search in $x$, $y$, and intensity space for the associated PSF reference images. The best-fit parameters for the three-dimensional grid search is determined by minimizing the diffraction spike residuals after the subtraction of the reference star PSF from the target image. For more details on the PSF subtraction technique described above, see Section 3 and Figure 1 in \cite{Avsar2024}.

Finally, all PSF-subtracted target images are rotated into the "north up" and "east left" orientation before being median combined. The uncertainty maps are generated by taking the standard error of the mean for each pixel across all combined images. The SNR and number of images per pixel maps can be found in Figure \ref{fig:npix-SNR}.

\begin{figure*}
    \centering
    \includegraphics[width=\linewidth]{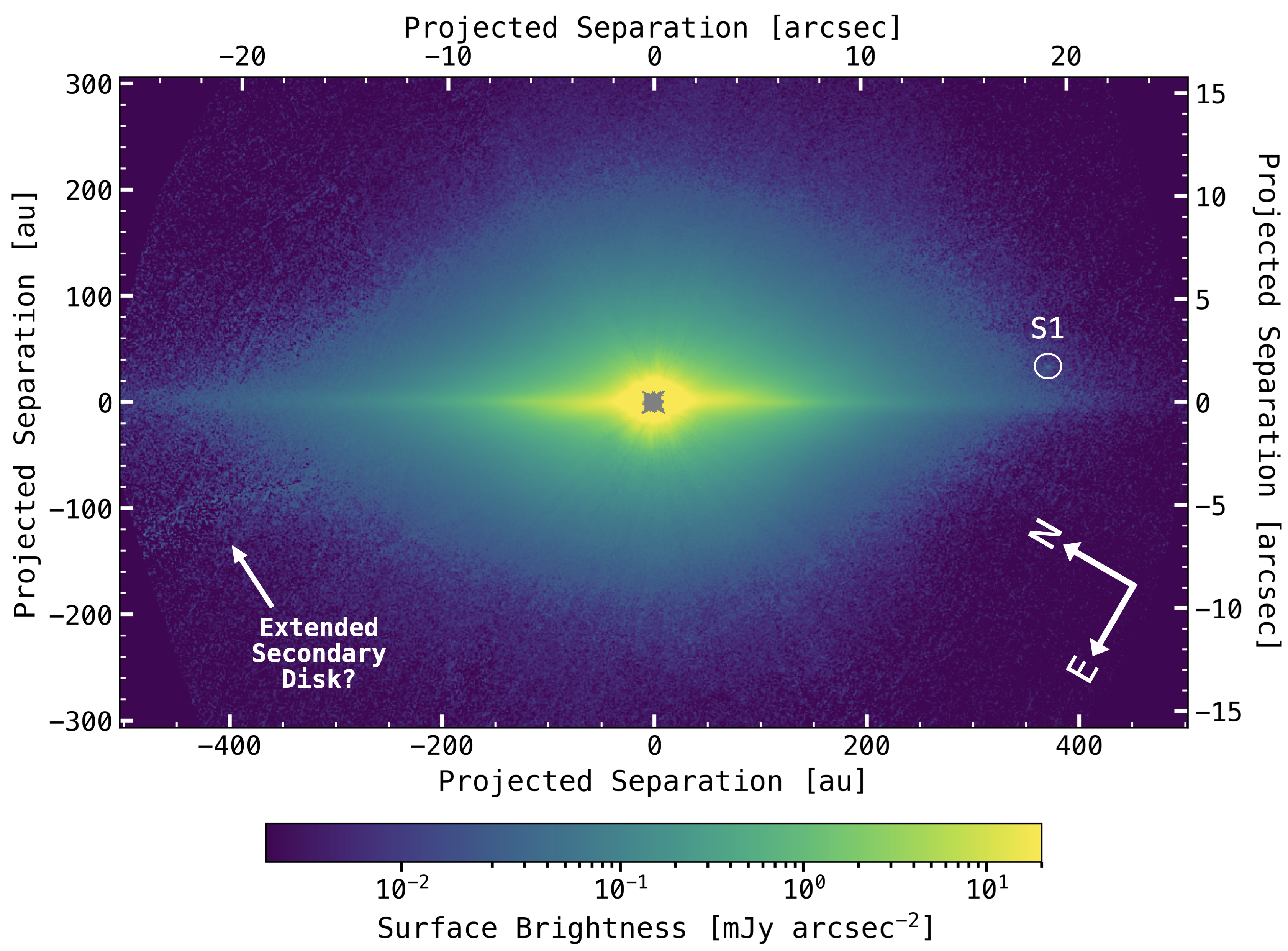}
    \caption{The $\beta$ Pictoris debris disk imaged with HST/STIS combining imaging from 2012 to 2025 on a logarithmic scale. The white circle indicates the S1 background source (Appendix \ref{sec:S1-astrometry}). The white arrow indicates the feature extending past the halo, possibly being the scattered light component of the extended secondary disk seen in \cite{Rebollido2024}.}
    \label{fig:main-image}
\end{figure*}

\begin{figure*}
    \centering
    \includegraphics[width=\linewidth]{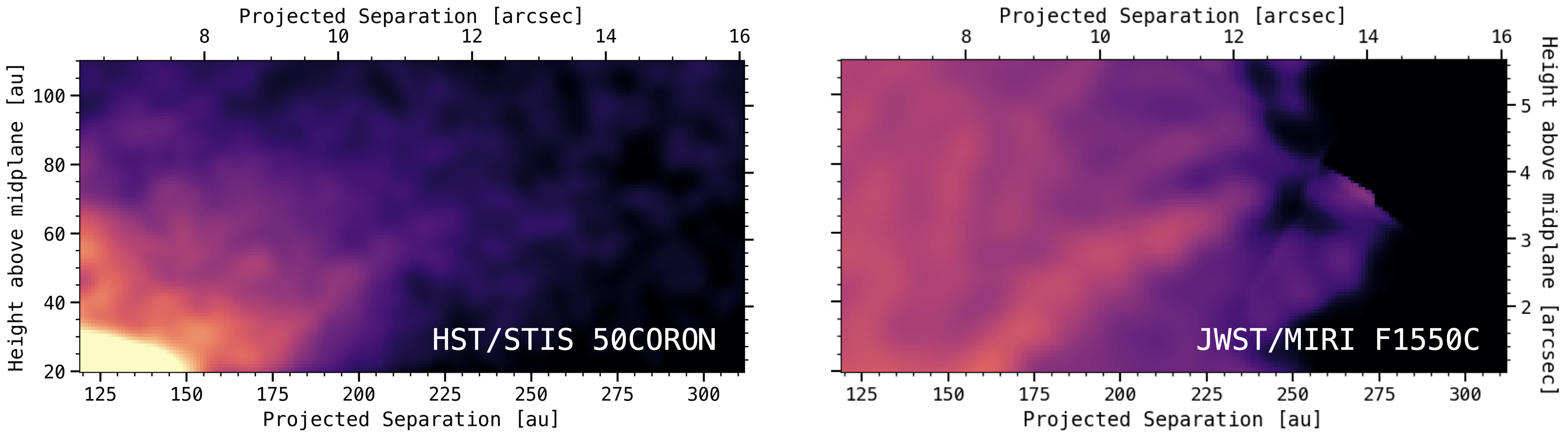}
    \caption{\textit{Left:} Zoomed in HST/STIS image of $\beta$ Pic from this work, indicating the detection of the scattered light component of the Cat's Tail with the white arrow. The shown image has undergone halo subtraction (see Figure \ref{fig:halo_asymmetries}), unsharp masking, and Gaussian smoothing to highlight the feature. \textit{Right:} Zoomed in JWST/MIRI F1550C image of $\beta$ Pic from \cite{Rebollido2024} highlighting the location of the Cat's Tail in comparison to the feature seen with HST/STIS.}
    \label{fig:cats-tail-stis-miri}
\end{figure*}

\begin{figure*}
    \centering
    \includegraphics[width=\linewidth]{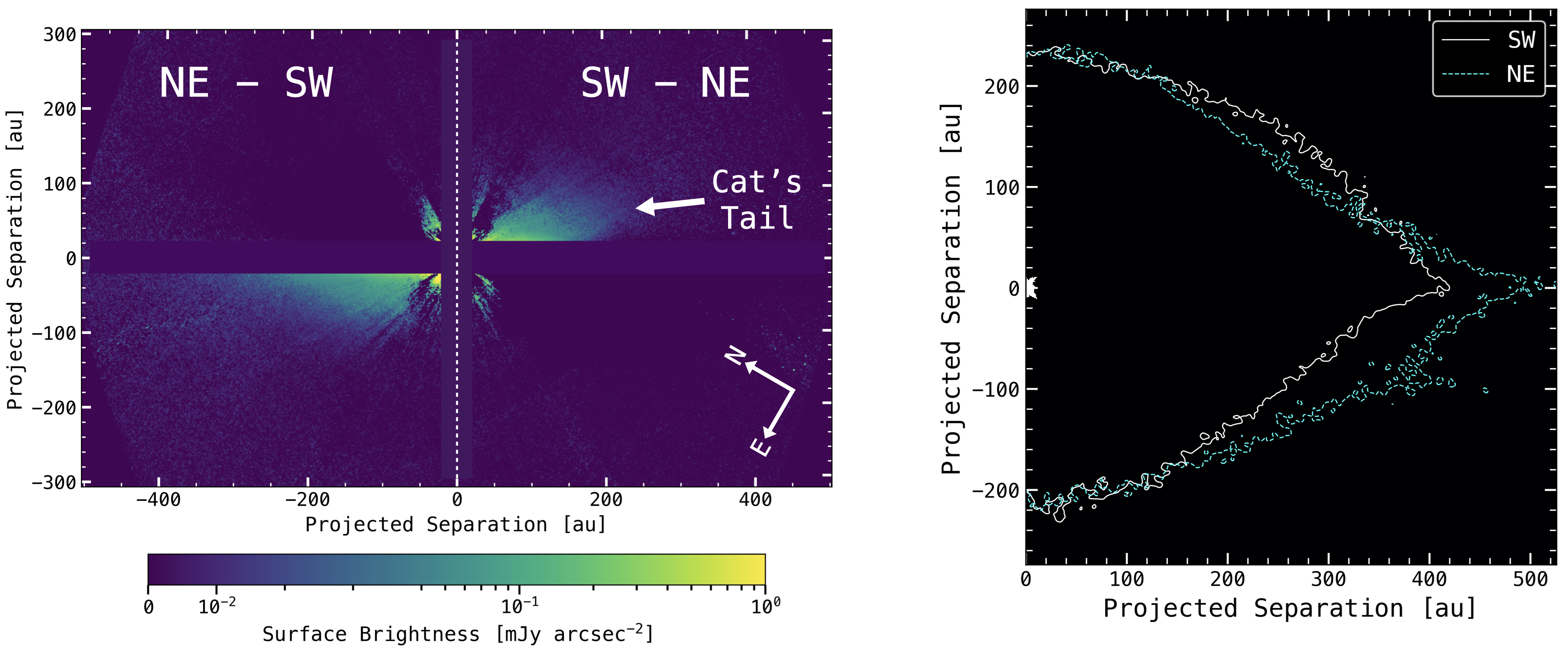}
    \caption{\textit{Left:} Image showing the NE and SW sides of the disk subtracted from each other with respect to the axis perpendicular to the midplane (denoted by the dashed white line). The resulting image reveals a non-axisymmetric halo and clear surface brightness asymmetry previously characterized as the butterfly asymmetry \citep[][]{Kalas&Jewitt1995}. With the halo above the SW midplane of the disk having excess surface brightness compared to above the NE midplane, and the halo below NE midplane having excess surface brightness compared to below the SW midplane. The white arrow points to the signal from the scattered light component of the Cat's Tail. \textit{Right:} Contour plot of the NE and SW halo plotted over one another. The image was smoothed using of Gaussian kernel of two pixels to reduce the noise of the low surface brightness halo. The contour traces regions of 0.01 mJy arcsec$^{-2}$ as that is what traces the outer edges of the halo. The contours depict a clear morphological asymmetry in addition to the surface brightness asymmetry on the left panel. The halo above the SW midplane is more vertically extended past $\sim$ 130 au. Below the midplane, the SW halo is more vertically extended compared to the NE side, which also starts at $\sim$ 130 au.}

    \label{fig:halo_asymmetries}
\end{figure*}

\section{Observational Results}\label{Observational-Results}

With the new observing configuration designed to detect the outer disk of $\beta$ Pic, we are able to detect the disk out to $\sim25$ arcseconds on each side of the disk for the first time with HST/STIS. This translates to a projected separation of approximately 500 au. 

The surface brightness of $\beta$ Pic varies by three orders-of-magnitude in our observations (see Figures \ref{fig:main-image} and \ref{fig:SB-profile}), from > 100 mJy arcsec$^{-2}$ in the inner disk ($<$ 50 au) to  $< $ 0.1 mJy arcsec$^{-2}$ in the outer reaches of the disk ($>$ 400 au). Additionally, we achieve a high signal-to-noise ratio (SNR) throughout large swathes of the disk (Figure \ref{fig:npix-SNR}). From a projected separation of 50 to 200 au, we consistently achieve an SNR of over 100, reaching a maximum SNR of $>$ 400 at $\sim 100$ au. At separations $>$ 200 au, a combination of lower dust density and the $r^{-2}$ falloff of surface brightness in scattered light observations leads to consistently decreasing SNR, with outer reaches of the disk detected with an SNR of $<$ 10.

We highlight that previous observations of $\beta$ Pic with HST/STIS only imaged out the disk out to 10 arcseconds or 200 au \citep{Apai2015,Avsar2024}. The region of the disk that is > 200 au has been imaged for the first time with HST/STIS with HST-GO-17456 (PI: Wagner).

\subsection{Detection of the Scattered Light Component of the Cat's Tail with HST/STIS}\label{cats-tail-section}

We report the first detection of the scattered light component of the Cat's Tail substructure in deep coronagraphic imaging of the $\beta$ Pic debris disk with HST/STIS (Figure \ref{fig:cats-tail-stis-miri}). The flux of the feature is low compared to the discovery observations with JWST/MIRI ($\sim200-1000\times$ lower than MIRI, see Section \ref{flux-constraint-section} for more details). The Cat's Tail substructure is not easily visible in the final median-combined image due to the high surface brightness of dust halo seen with HST/STIS (Figure \ref{fig:main-image}).

To reveal the Cat's Tail, we first subtract the surface brightness from the outer dust halo by subtracting the NE halo from the SW side of the halo where the Cat's Tail feature is located. This removes a majority of the flux from the SW halo and reveals the outline of the Cat's Tail (see left panel of Figure \ref{fig:halo_asymmetries}). We then apply an unsharp mask to further remove the smooth residual halo, followed by Gaussian smoothing ($\sigma = 3$) to suppress pixel-to-pixel noise. The final version highlighting the signal from the optical component of the Cat's Tail is shown in Figure \ref{fig:cats-tail-stis-miri}.

\begin{figure*}
    \centering
    \includegraphics[width=\linewidth]{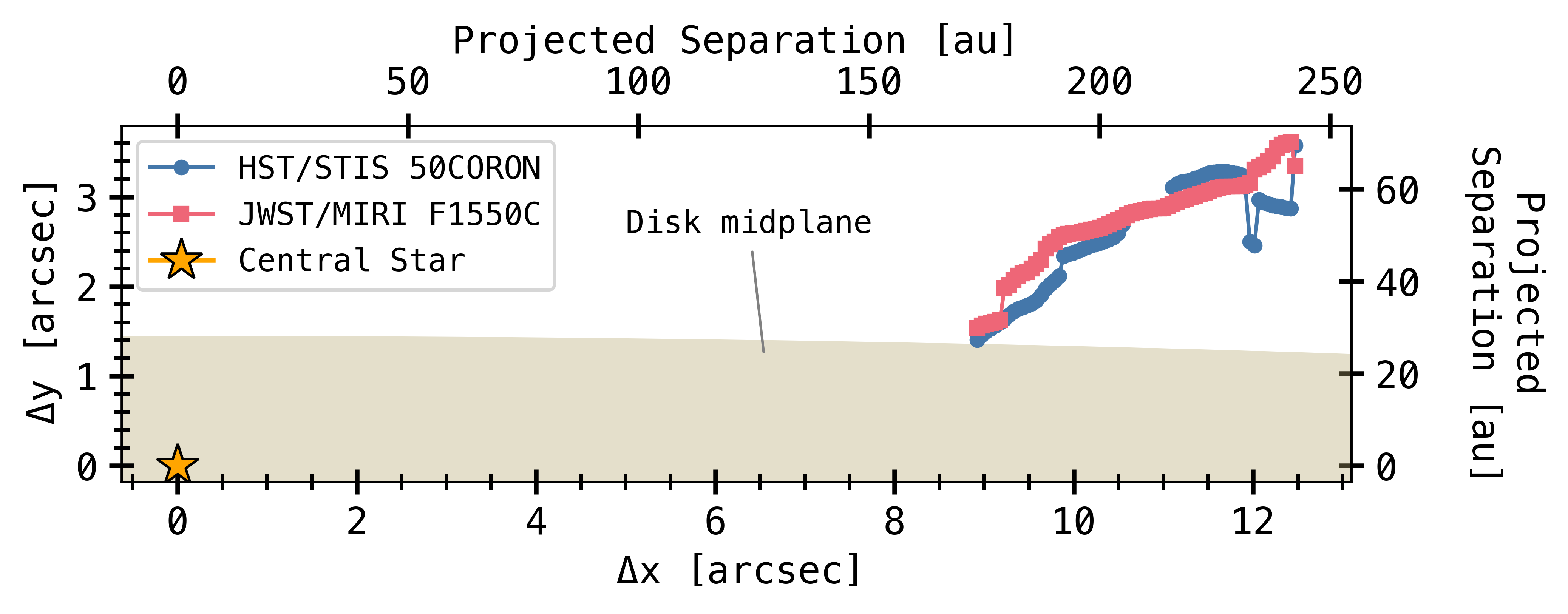}
    \caption{Comparison of the spine of the Cat's Tail in HST/STIS versus JWST/MIRI F1550C. Each point marks the brightest pixel in a horizontal cut perpendicular to the tail. We find that the morphology of the Cat's Tail with STIS matches the MIRI morphology well but not perfectly, especially towards the tip of the tail. Since the tip is the furthest from the central star, the scattered light component of the tail becomes indistinguishable from the background at the largest separations, while the thermal component is still detectable, causing the observed divergence.}
    \label{fig:cats_tail_trace}
\end{figure*}

\begin{figure}
    \centering
    \includegraphics[width=\linewidth]{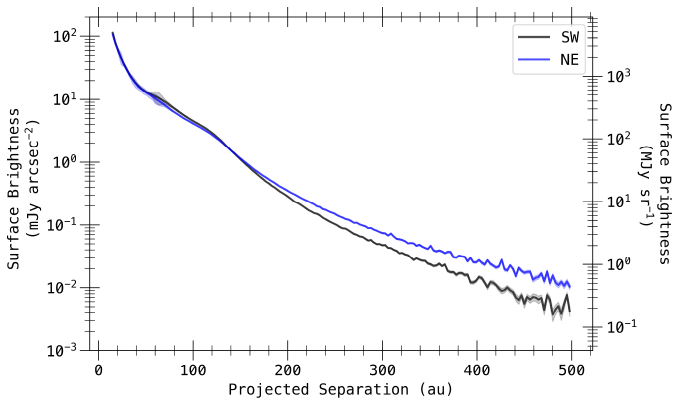}
    \caption{Midplane radial surface brightness profiles of the $\beta$ Pic debris disk, with the envelopes indicating the 1$\sigma$ uncertainty. We perform measurements from 20 to 500 au in projected separation.}
    \label{fig:SB-profile}
\end{figure}

The feature detected with HST/STIS matches the morphology of the Cat's Tail seen with JWST/MIRI well, making it unlikely that the feature seen with HST/STIS is noise or an image processing artifact (see Appendix \ref{sec:jackknife} and Figure \ref{fig:jackknife}). However, the morphological match is not perfect (see Figure \ref{fig:cats_tail_trace}), with the spines diverging no more than 0.3 arcseconds from projected separations of 9 to 12 arcseconds. We find that the upward curvature of the Cat's Tail seen with MIRI is not detected with STIS. More specifically, we do not detect the tip of the MIRI Cat's Tail where it is seen to curve up at $\sim12$ arcseconds, becoming indistinguishable from the background (see Section \ref{sec:truncation} for more detail). This is confirmed both visually in Figure \ref{fig:cats-tail-stis-miri} and by the divergence of the spine seen with STIS from that of MIRI past 12 arcseconds in Figure \ref{fig:cats_tail_trace}. This places the maximum projected stellocentric separation of the Cat's Tail seen with STIS at $\sim230$ au. The Cat's Tail is detected with an SNR of $\sim10$ and surface brightness of $\sim0.03$ mJy arcsec$^{-2}$ above the halo background at the base, and an SNR of $\sim5$ and surface brightness of $\sim0.01$ mJy arcsec$^{-2}$ above the halo background at the tip. In the coming sections, we will further characterize the flux of the optical component of the Cat's Tail relative to that seen with MIRI and model the properties of the dust grain population within the Cat's Tail substructure.

\subsection{Potential Low SNR Detection of Extended Secondary Disk in STIS Observations}

\begin{figure*}
    \centering
    \includegraphics[width=\linewidth]{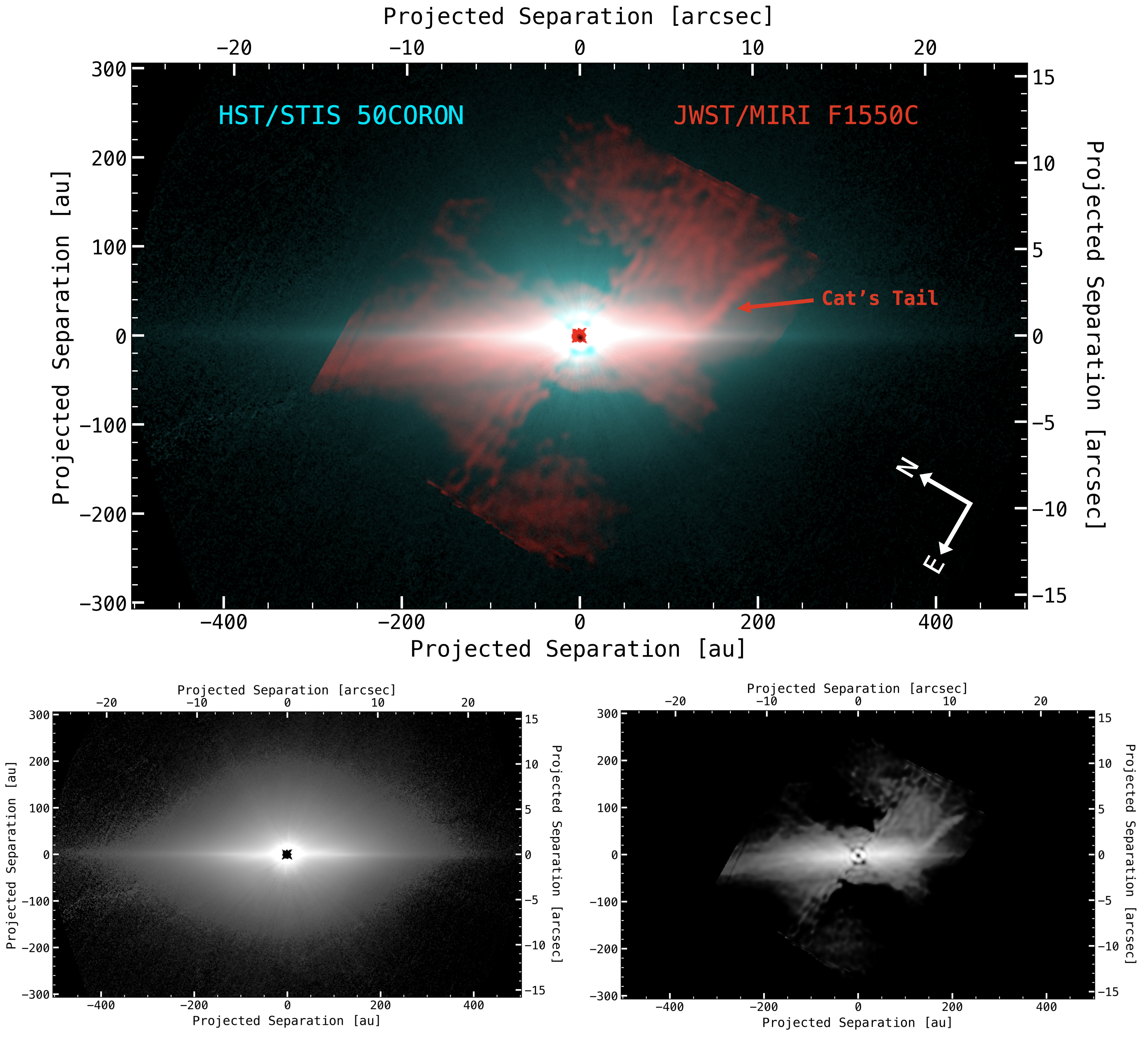}
    \caption{RGB image of the HST/STIS image presented in this work overlaid with the deconvolved F1550C image from \cite{Rebollido2024}. The bottom two images show a grayscale version of each dataset in the RGB image. We highlight the smooth/axisymmetric halo in the HST/STIS image in contrast to the complex substructures seen in the F1550C image. }
    \label{fig:rgb}
\end{figure*}

We note the detection of a low SNR feature below the NE midplane that appears to extend past the dust halo (see Figures \ref{fig:main-image} and \ref{fig:npix-SNR}). The feature is located near the same location as the extended secondary disk seen in the MIRI data \citep[][]{Rebollido2024}. Since the field of view of the MIRI observations does not extend past the halo observed with STIS, it is difficult to determine whether the signal seen with STIS is offset from the extended secondary disk or whether the extended secondary disk would be coincident with the feature seen with STIS (see Figure \ref{fig:rgb}).The low SNR nature of the signal and the fact that the structure may point at the central star makes it so that we cannot rule out that the observed structure is a PSF artifact. However, deeper observations of the $\beta$ Pic outer disk as described in Table \ref{tab:obs-2} should be able to determine whether this structure is the scattered light component of the extended secondary disk seen with MIRI or a PSF artifact.

\bigskip

\subsection{Surface Brightness Asymmetry and Warped Morphology in the Outer Dust Halo}\label{convave-convex-section}

The new wide field-of-view observations of the disk from 2023 to 2025 allow for the characterization of the scattered light debris halo around $\beta$ Pic at wide separations and high angular resolution that was not possible in previous observations from space or the ground. Previous ground-based, wide field-of-view observations by \cite{Kalas&Jewitt1995} covered similar separations but with an effective angular resolution of 1.4 arcseconds, while the presented HST/STIS observations have a resolution 0.07 arcseconds. This is a roughly 20$\times$ increase in angular resolution. We define the halo of the $\beta$ Pic debris disk as signal from regions above or below the midplane of the disk. The larger field of view observations reveal that the halo of the dust surrounding $\beta$ Pic is not axisymmetric, with a warp in the morphology of the halo in addition to a clear surface brightness asymmetry between the NE and SW sides of the halo (see Figure \ref{fig:halo_asymmetries}). 

To characterize the surface brightness asymmetry of the halo, we mask the midplane of the disk and subtract each side of the disk from the opposite side with respect to the axis perpendicular to the disk midplane (see left panel of Figure \ref{fig:halo_asymmetries}). This subtraction reveals the surface brightness asymmetry in the halo. We find the halo above the midplane on the SW side of the disk having a larger surface brightness than the halo above the NE side, and the halo below NE side of the disk having a larger surface brightness than the halo below the SW side. The observed surface brightness asymmetry is on the order of 0.1 mJy arcsec$^2$ at the base of the halo and on the order of 0.01 mJy arcsec$^2$ where the halo is > 100 au above the midplane.

To characterize the morphological asymmetry of the halo, we start by applying a two pixel width Gaussian filter to the image of $\beta$ Pic to reduce the noise of the low surface brightness halo outer border. We then overplot contours of 0.01 mJy arcsec$^2$, which represent the outer edge of the halo, to reveal the asymmetry in the outer halo morphology. Above/north the midplane, we find the SW side of the halo exhibiting a convex morphology, and the NE side exhibits a concave morphology, leading to the SW side to be more vertically extended than the NE side. The opposite is observed in the halo below/south the midplane, where the NE halo exhibits a convex morphology compared to the SW exhibiting a concave morphology. These morphological asymmetries start at projected separations of $\gtrsim$130 au.

\bigskip

\subsection{S1 Point Source}

In our final median combined images, we detect only a single point source, which we call S1 (Figure \ref{fig:main-image}). S1 is located at a projected separation of $\sim$ 400 au and is inclined relative to the disk midplane by $\sim4^\circ$. We use aperture photometry to measure the brightness of the S1 source = 1.58 $\pm$ 0.03 $\mu$Jy in the STIS white light bandpass. In Appendix \ref{sec:S1-astrometry}, we perform an astrometric analysis demonstrating that S1 is likely a background source.

\begin{figure*}
    \centering
    \includegraphics[width=\linewidth]{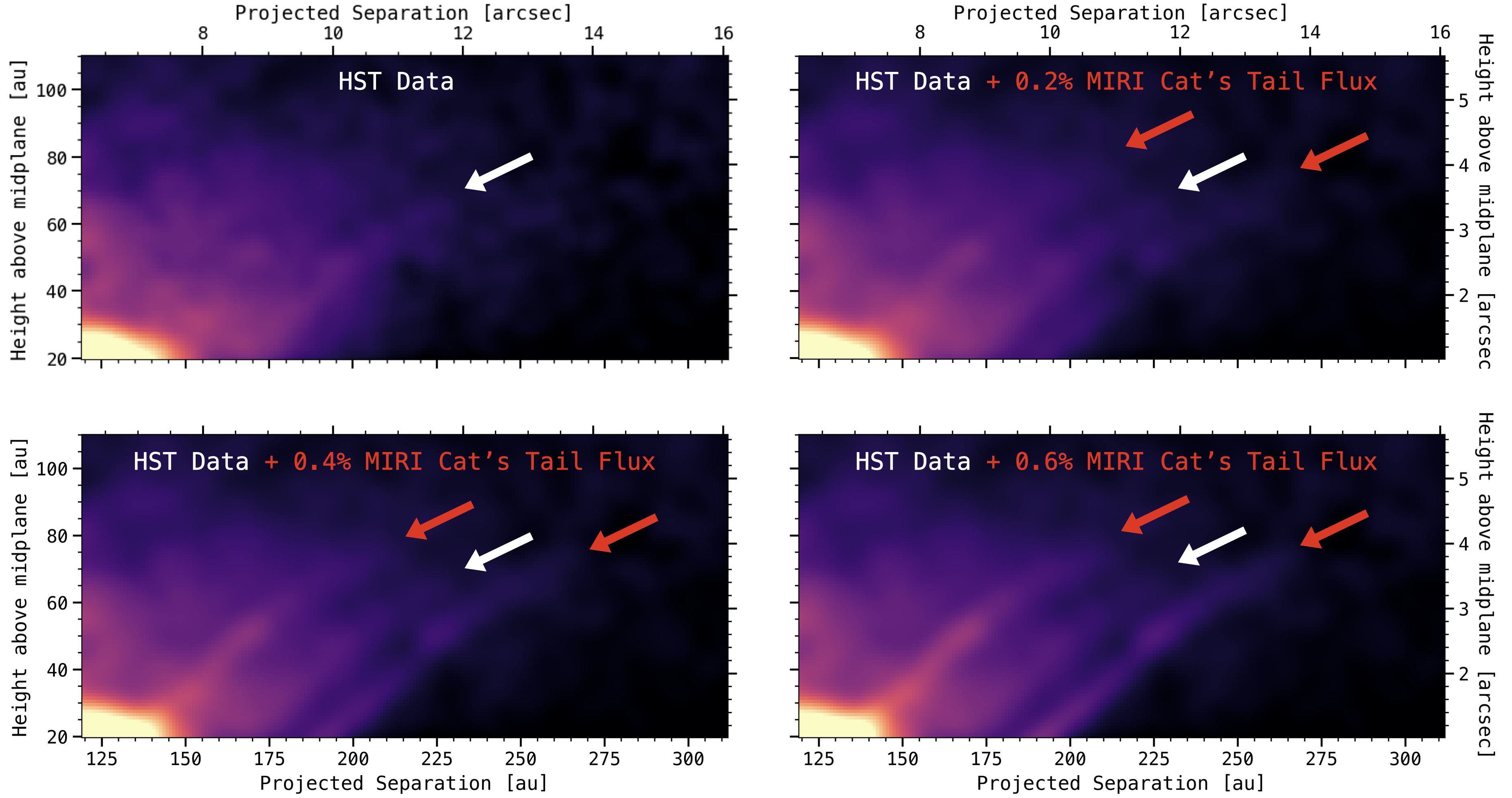}
    \caption{Gallery of injections of two Cat's Tail models into HST/STIS observations of $\beta$ Pic. To best sample the background noise, we inject two Cat's Tail models in each image, one to the left and one to the right of the detected structure. We show injections of three different Cat's Tail scattered light fluxes, ranging  from 0.2$\%$ to $0.6\%$ the observed MIRI flux at 15.5 $\mu$m. The white arrow indicates the Cat's Tail structure seen in the data, while the red arrows point to the injected features.}
    \label{fig:cats_tail_injections}
\end{figure*}

\begin{figure}
    \centering
    \includegraphics[width=\linewidth]{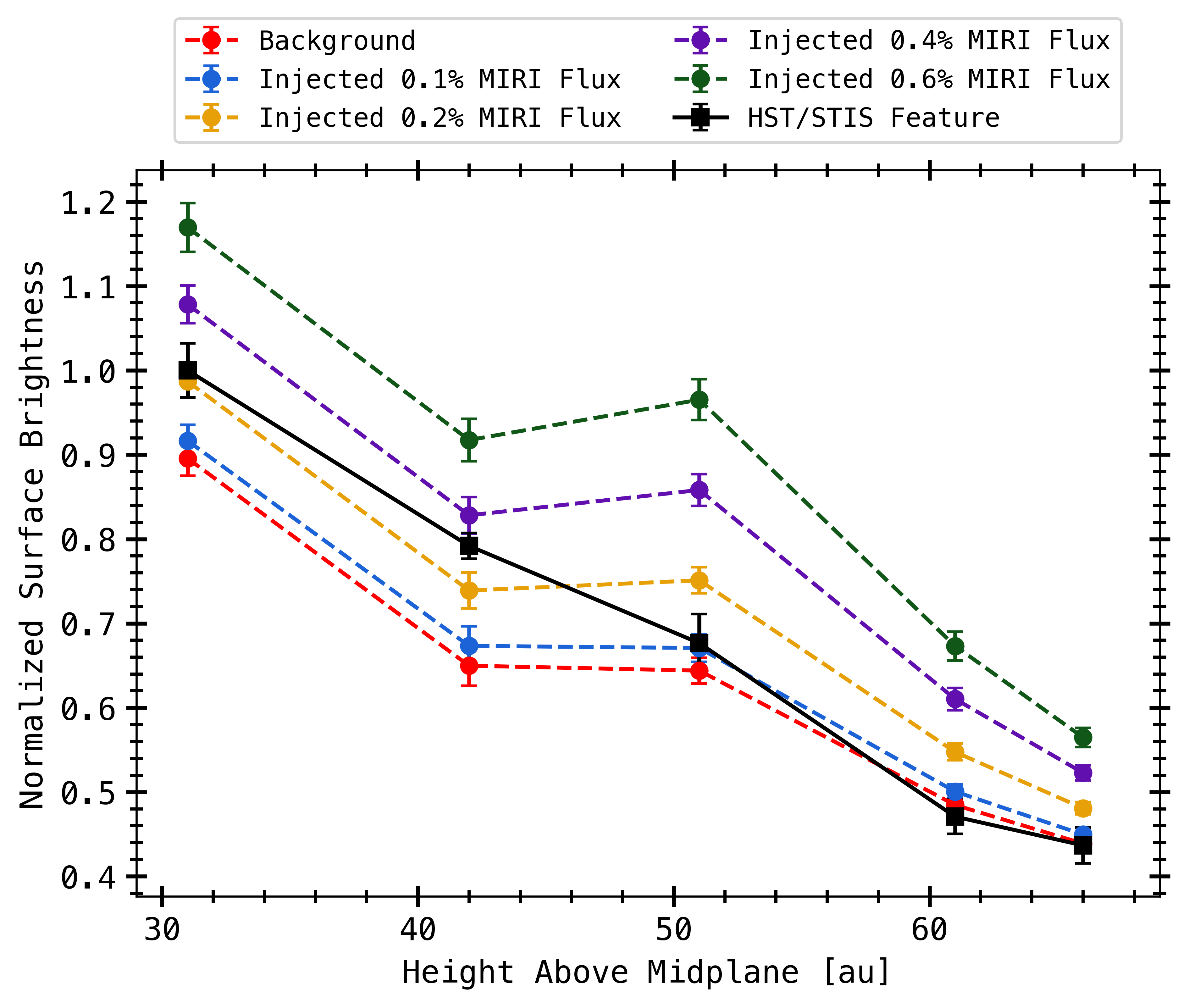}
    \caption{Normalized surface brightness measurements of the detected Cat's Tail feature, injected features, and the background of region the model Cat's Tails were injected. Since two Cat's Tail features were injected in each image to best sample the background noise, the data points show the mean of the two apertures for the same height above the midplane. We find that the detected feature has a flux of $0.1\%-0.4\%$ the MIRI flux, with the feature becoming indistinguishable from the background beyond $\sim55$ au above the midplane. }
    \label{fig:SB_tails}
\end{figure}

\section{Injection/Recovery Analysis of the Cat's Tail Morphology and Surface Brightness }\label{Injection-Recovery}

With the discovery of the visible/near-infrared component of the Cat's Tail substructure first seen with JWST/MIRI, we seek to constrain the flux of the Cat's Tail in scattered light versus thermal emission in the mid-IR. This is done with the ultimate goal of characterizing the dust grains responsible for the observed substructure. In this section we will outline the Cat's Tail model developed and applied to the HST/STIS observations.

\subsection{Creating an HST/STIS Resolution Cat's Tail Model}

We start by re-projecting the MIRI F1550C image onto the STIS pixel grid, resampling from 0.11 arcsec/pixel to 0.05 arcsec/pixel. We then create a model of the projected Cat's Tail, which when convolved to the MIRI F1550C resolution (FWHM $\sim$ 0.5 arcsec) visually matches the size of the observed feature in the mid-IR. We adjusted the width of the thin model until the thin model that is convolved with MIRI F1550C resolution matched the width of the Cat's Tail in the original MIRI image. That results in a thin Cat's Tail model width of between 6-8 pixels throughout the tail.

We first must subtract the background surface brightness in the region of the Cat's Tail in the MIRI image from the dust halo/nebulosity. To remove the background, we mask the Cat's Tail in the F1550C image and assign Gaussian weights to pixels adjacent to the Cat's Tail, with the weight decreasing as a function of distance. That is done for each pixel in the mask and combined into a background map. We then subtract the background map from the Cat's Tail in the MIRI image to remove the background surface brightness from the Cat's Tail in the MIRI image.

For accurate comparison between STIS and MIRI observations, it is important that the thin Cat's Tail model conserves the flux and surface brightness gradient of the MIRI Cat's Tail. We center the thin model over the F1550C Cat's Tail and assign the overlapping pixel values from the Cat's Tail to the thin model. We then take the flux ratio of the sum of the entire MIRI Cat's Tail and the region of the MIRI Cat's Tail that overlapped with the thin model. We multiply the thin Cat's Tail by the flux ratio to ensure flux conservation between the MIRI Cat's Tail and the thin Cat's Tail model. Finally, the thin Cat's Tail model is convolved with a Gaussian equivalent of the HST/STIS 50CCD PSF (FWHM $\sim$ 0.07 arcsec).

\subsection{Injection of the Cat's Tail Feature into the HST/STIS Image of $\beta$ Pic}

We first convert each individual image in the presented STIS observations of $\beta$ Pic from mJy arcsec$^{-2}$ to MJy sr$^{-1}$ to match the units of the MIRI observations. Once the Cat's Tail model for HST/STIS observations is created, it is injected into each of the 127 images with photon noise, re-reduced and median combined. This is done for the scattered light Cat's Tail with $0.1\%$ to $0.6\%$ the flux of the Cat's Tail seen with MIRI F1550C (see Figures \ref{fig:cats_tail_injections} and \ref{fig:SB_tails}). To best sample to background noise of the detected scattered light Cat's Tail, two Cat's Tail models are injected 20 au to the left and the right of the observed feature.

For each set of Cat's Tail injections, we place five apertures along the structure at equal separations above the midplane to measure the average surface brightness that could then be compared to the detected feature. We additionally measure the background surface brightness in the injection regions to show that the detected structure is above the background dust halo values. Since we inject two offset features for each flux value, we take the mean of the two apertures at equal heights above the midplane.

\subsection{Constraints on the Cat's Tail Surface Brightness in Scattered Light}\label{flux-constraint-section}

The surface brightness measurements of the detected feature in HST/STIS observations, the mean of the injected Cat's Tail models, and the background can be found in Figure \ref{fig:SB_tails}. We find that the observed flux of the scattered light component of the Cat's Tail with HST/STIS is between $0.1\%$ and $0.4\%$ the flux of the Cat's Tail seen with MIRI F1550C. We find that the base of the observed tail is closer to between $0.2\%$ and $0.4\%$ from 30 to 40 au above the midplane, but is better matched at the $0.1\%$ level from 40 to 50 au above the midplane, leading to our range of between $0.1\%$ and $0.4\%$ the flux of the Cat's Tail seen with MIRI F1550C. In our data, at separation $>55$ au above the midplane, the surface brightness of the Cat's Tail becomes indistinguishable from the mean background surface brightness. This is consistent with the visual comparison of the HST/STIS and JWST Cat's Tail features (Figure \ref{fig:cats-tail-stis-miri}), where the Cat's Tail appears more vertically extended with MIRI than STIS.

\section{Modeling of Grain Properties in the Cat's Tail}\label{Grain-Properties}

In this section, we will outline the model we use to place constraints on the properties of the dust grains of the Cat's Tail. Specifically, we will fit dust grain properties that are consistent with our scattered light observations.

We determine the grain properties by weighting the composition of the dust grain between organic refractory material, astronomical silicates, and vacuum/porosity. The properties of the grains are determined by Mie theory. We use optical constants for organic refractory material from \cite{Li&Greenberg1997} and astronomical silicates from \cite{Draine1984} (see Appendix \ref{pyox-olivine} and Figure \ref{fig:other-mcmc} for the implementation of optical constants for pyroxene and olivine instead of astronomical silicates). These optical constants for a mixture of materials are combined using the Bruggeman mixing rule, which allows for varying compositions and porosity. The combined optical properties are then used to determine the scattering efficiency factor ($Q_{scat}$), absorption efficiency factor ($Q_{abs}$), and radiation pressure efficiency factor ($Q_{pr}$). $Q_{pr}$ is used in determining the ratio of the force of radiation pressure to the force of gravity ($\beta$) felt by the dust grain, while $Q_{scat}$ and $Q_{abs}$ are used in determining the flux density from grains in the scattering and thermal emission regimes, respectively.

For the central star in the $\beta$ Pic system, we assume a photospheric temperature of 8054 K \citep[][]{Baburaj2026}, a stellar radius of $R_\star$ = 1.6 $R_{\odot}$ \citep[][]{Stassun2019}, a mass of $M_\star$ = 1.8 $M_{\odot}$ \citep[][]{Nowak2020}, and at a distance of $d_{\star}=19.8$ pc \citep[][]{Gaia2021}.

\subsection{Calculating the Flux Ratio Between Observations}

 To fit the scattered light flux ratio range (0.1--0.4$\%$), we calculate the band-averaged flux density of a dust grain in the scattered light regime with the following: 

\begin{equation}
\resizebox{0.97\columnwidth}{!}{$
\displaystyle \mathcal{F}_{\mathrm{STIS}} = \left(\frac{R_{\star}}{r}\right)^2 \left(\frac{\pi s}{d_{\star}}\right)^2 \frac{\displaystyle\int Q_{{scat}}(s,\lambda_S)\, B_\star(T_\star)\, \Phi(\theta,g(\lambda_S))\, t_S(\lambda_S)\, \lambda_S\, d\lambda_S}{c \displaystyle\int t_S(\lambda_S)/\lambda_S\, d\lambda_S}
$}
\end{equation}

where $s$ is the radius of the dust grain, $r$ is the true stellocentric distance of the grain, $c$ is the speed of light, $B_{\star}$ is the Planck function for $\beta$ Pic, $t_S$ is the STIS transmission curve at a given wavelength $\lambda_S$. $\Phi$ represents the Henyey-Greenstein scattering phase function for a given scattering angle, $\theta$, and $g$ is the asymmetry parameter and is calculated from the grain's optical constants. 

The scattering angle is defined as the angle between the vector along which the photon would have traveled if it had not encountered the dust grain and the vector along which the photon actually travels after being scattered. The Henyey-Greenstein scattering phase function is given by: 

\begin{equation}
\Phi(\theta, g) = \frac{1 - g^2}{4\pi\left(1 + g^2 - 2 g \cos\theta\right)^{3/2}}
\end{equation}

\begin{figure}
    \centering
    \includegraphics[width=\linewidth]{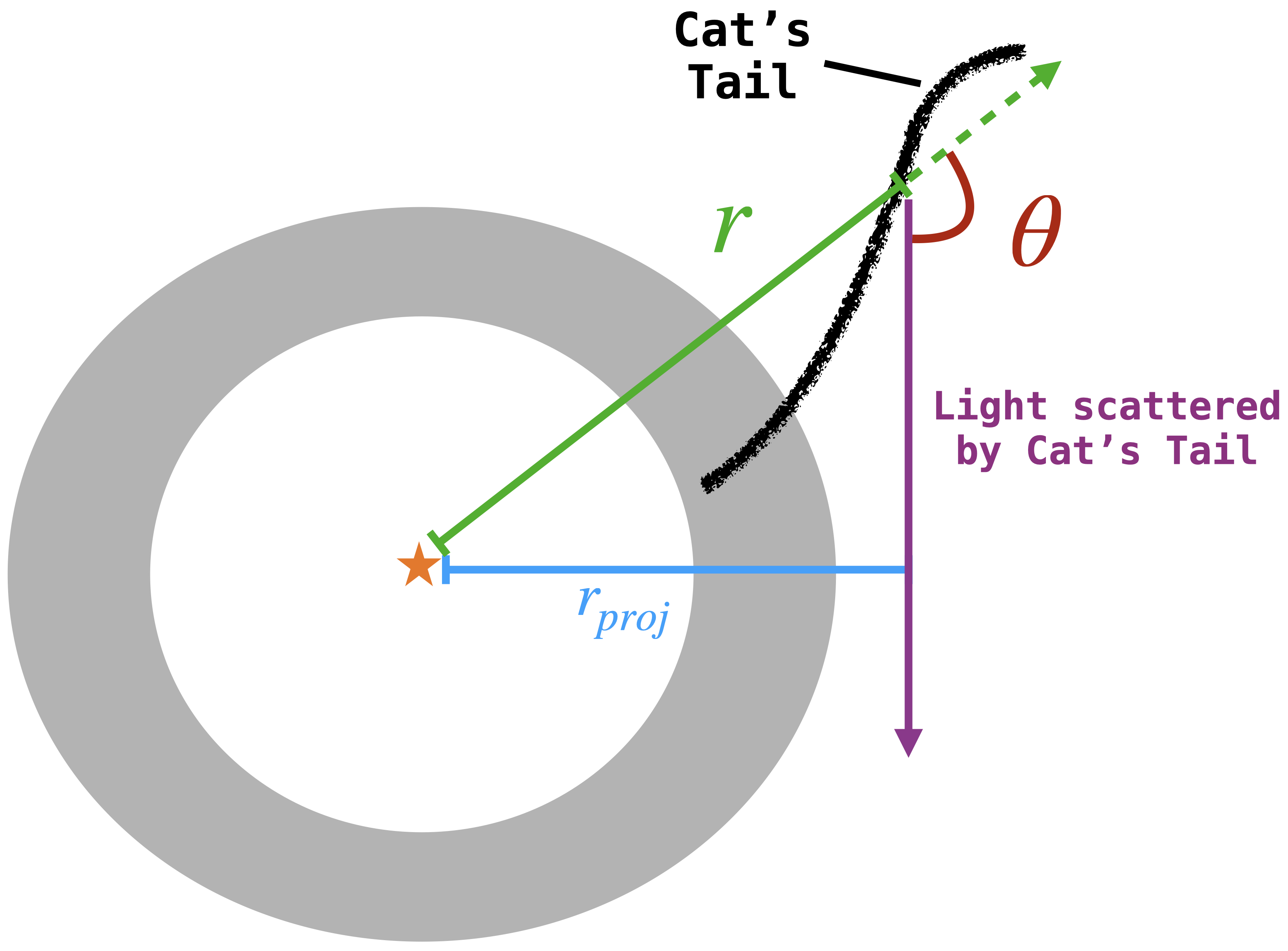}
    \caption{Face-on schematic of the $\beta$ Pic disk and the Cat's Tail substructure. The schematic highlights the relationship between the true separation ($r$), projected separation ($r_{proj}$), and the scattering angle of the dust in the Cat's Tail ($\theta$).}
    \label{fig:scattering_schematic}
\end{figure}

The true stellocentric distance of the grains in the Cat's Tail ($r$) is determined by a combination of the projected separation and the scattering angle. Since the projected separation of the Cat's Tail is known, the scattering angle and true separation of the Cat's Tail must be consistent with the projected separation of the feature. The relation between the true separation, projected separation, and scattering angle is given by: 

\begin{equation}
    r(r_{proj},\theta) = \frac{r_{proj}}{\sin(180^{\circ}-\theta)} =  \frac{r_{proj}}{\sin(\theta)}
\end{equation}

where $r_{proj}$ is the projected separation in au and $\theta$ is the scattering angle in degrees. We choose to anchor our model fitting to the midpoint of the Cat's Tail in projected separation, which is roughly 175 au (see Figure 9 in \cite{Rebollido2024}). A schematic demonstrating the scattering geometry can be found in Figure \ref{fig:scattering_schematic}.

The flux ratio is then determined for the grain in the mid-IR, specifically for the F1550C filter, given by:

\begin{equation}
\resizebox{0.97\columnwidth}{!}{$
\displaystyle \mathcal{F}_{\mathrm{F1550C}} = \pi  \left(\frac{s}{d_{\star}}\right)^2 \frac{\displaystyle\int Q_{\mathrm{abs}}(s,\lambda_M)\, B_{\mathrm{dust}}(T_{\mathrm{dust}},s)\, t_M(\lambda_M)\, \lambda_M\, d\lambda_M}{c \displaystyle\int t_M(\lambda_M)/\lambda_M\, d\lambda_M}
$}
\end{equation}

where  $Q_{abs}$ is the absorption efficiency factor, $B_{dust}$ is the Planck function for the dust grain, and $t_M$ is the F1550C transmission curve for a given wavelength, $\lambda_M$. Since the mid-IR flux is dominated by thermal emission rather than scattered light, it is assumed it emits isotropically. Furthermore, $T_{dust}$ is determined as the equilibrium temperature for a dust grain of radius $s$. 

We then find the dust grain size-averaged flux ratio between the STIS and MIRI F1550C observations using:

\begin{equation}\label{ratio}
 \frac{\mathcal{F}_{\mathrm{STIS}}}{\mathcal{F}_{\mathrm{F1550C}}}= \frac{\int_{s_{min}}^{s_{max}} n(s) \mathcal{F}_{scat}(s,\lambda_S) ds}{\int_{s_{min}}^{s_{max}} n(s) \mathcal{F}_{therm}(s,\lambda_M) ds}
\end{equation}

where $s_{min}$ and $s_{max}$ are the smaller and largest grain sizes in the Cat's Tail respectively. We choose an $s_{min}$ of 0.1 $\mu$m and an $s_{max}$ of 100 $\mu$m. As the Cat's Tail is thought to be populated by unbound grains, our choice of $s_{max}$ is motivated from \cite{Rebollido2024}. Figure 14 in \cite{Rebollido2024} shows that highly porous organic refractory grains as large as 100 $\mu$m can still be in the blow-out size regime ($\beta>0.5$), getting removed from the system by radiation pressure. $n(s)$ is the size distribution of dust grains of radius $s$. We choose to use a steeper-than-Dohnanyi size distribution where $n(s) \propto s^{-3.8}$ for a collisional cascade not yet in steady state. 

\subsection{Model Fitting of Dust Grain Properties}\label{model-fitting-section}

We fit the dust grain properties using {\tt\string emcee}, an Affine Invariant Markov Chain Monte Carlo (MCMC) Ensemble Sampler \citep{emcee1,emcee2}. In our model, we vary dust grain properties to be consistent with the scattered light component being $0.1-0.4\%$ the MIRI F1550C flux (0.001 < $\frac{\mathcal{F}_{\mathrm{STIS}}}{\mathcal{F}_{\mathrm{F1550C}}}$ $<0.004$). Our model includes three free parameters, the silicate fraction of the dust ($f_{sil}$), porosity of the dust grain ($P$), and the scattering angle of dust grains in the Cat's Tail relative to the observer $(\theta)$. The fraction of organic refractory material is determined by $f_{org} =1-f_{sil}$. In addition to the STIS to MIRI F1550C flux ratio, the model is informed by the flux ratio of the Cat's Tail between the F1550C and F2300C filters to be between 0.5 and 0.8 (see Figure 17 in \citealt{Rebollido2024}). 

From the best fit parameters, we will also derive the true separation of the Cat's Tail midpoint ($r$) and the dust temperature. The dust temperature will be incorporated by calculating size-averaged dust temperature for the grains in our model by:

\begin{equation}
    <T_{dust}> = \frac{\int_{s_{min}}^{s_{max}} T_{dust}(s)  s^2 n(s) ds}{\int_{s_{min}}^{s_{max}} s^2 n(s) ds}
\end{equation}

Additionally, our model will calculate $\beta$ for the largest grain present in the sample so that $\beta(s_{max}) \sim 10$. Since $\beta$ generally decreases at larger grain sizes, this ensures that all grains present in the population have a $\beta$ consistent with \cite{Rebollido2024}. Finally, the model will calculate the flux ratio between the F1550C and F2300C filters by adapting Equation \ref{ratio} for two filters in the thermal emission regime.

\subsection{Modeling Results}\label{model-results-section}

\begin{figure*}
    \centering
    \includegraphics[width=0.9\linewidth]{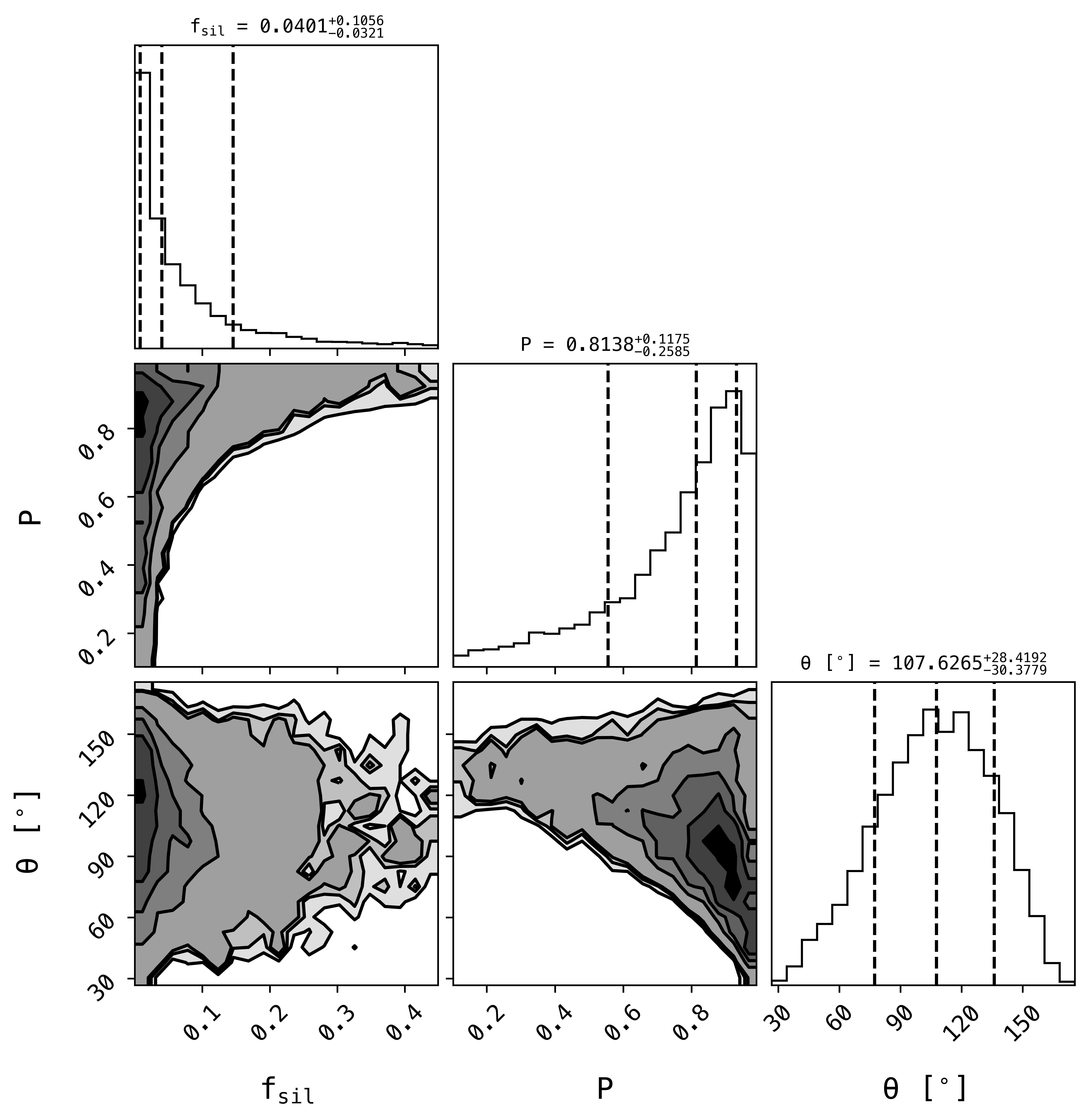}
    \caption{The posterior distribution of the {\tt\string emcee} chain of the three dust grain parameters of the Cat's Tail. The model is fit with 32 walkers, 10000 steps, and 1000 burn-in steps. The dashed lines indicate the 16th, 50th, and 84th percentiles, respectively. From these results, we conclude that the dust in the Cat's Tail is likely highly porous, organics rich, and located at or behind the plane of the sky relative to the observer.}
    \label{fig:mcmc}
\end{figure*}

We find that our STIS constraint of $0.1-0.4\%$ of the MIRI F1550C flux is best-fit by small, porous grains composed of primarily organic refractory material. The results of the modeling are shown in Figure \ref{fig:mcmc} and Table \ref{tab:disk_properties}.

Specifically, we find that the fraction of solid material in astronomical silicates in our fit is $1-15\%$, implying the fraction of solid material in organic refractory material is $85-99\%$. The dust in the Cat's Tail has a porosity of 0.55 to 0.93. Finally, we find that the scattering angle of the dust grains ranges from $\sim77^{\circ}-135^{\circ}$. From the best-fit parameters, we derive additional dust grain properties. We find a size averaged dust temperature ($<T_{dust}>$) of $\sim$ 120 -- 145 K and a midpoint true stellocentric separation ($r$) of 170--250 au.

\subsubsection{Impact of Dust Composition and Scattering Angle}

\begin{figure}
    \centering
    \includegraphics[width=\linewidth]{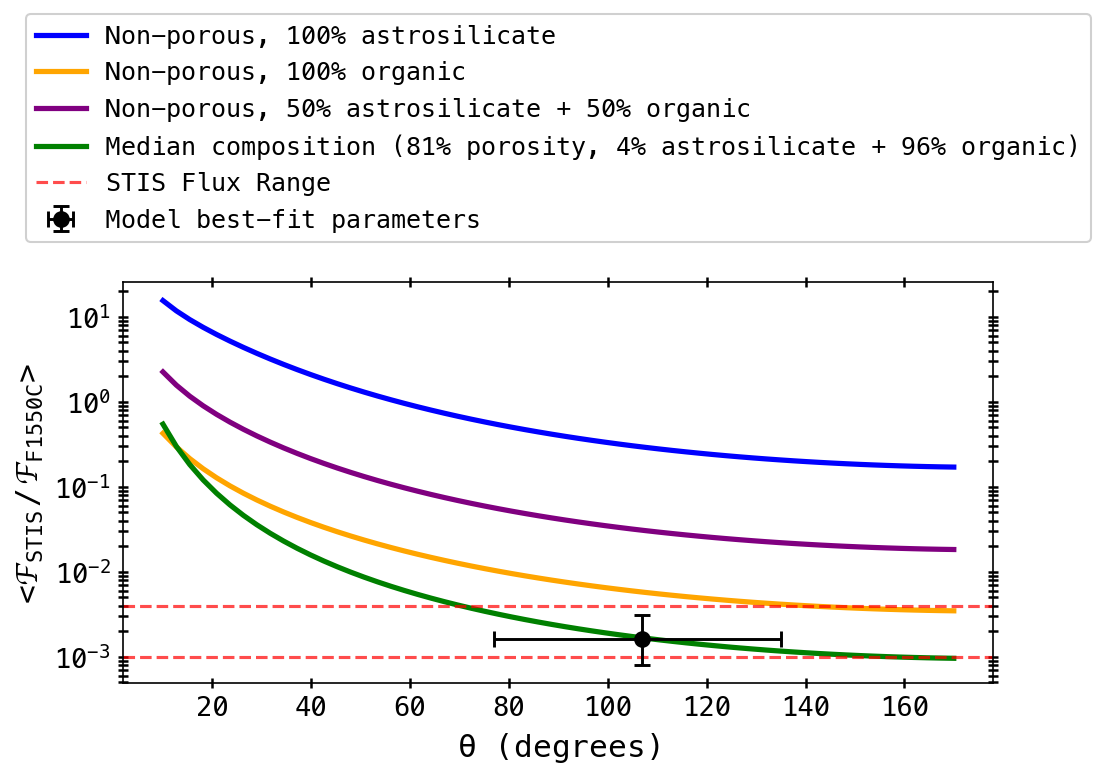}
    \caption{The STIS to F1550C flux ratio of dust grains of varying composition and porosity as a function of scattering angle at a median true separation of 184 au. We find a Cat's Tail composed largely of astrosilicates would be 0.1-10 times brighter at STIS wavelengths compared to F1550C. We find that, with constant composition and porosity, the Cat's Tail flux at STIS wavelengths can change by roughly three orders-of-magnitude as a function of the feature's scattering angles. The possible scattering angles given the feature's geometry are indicated by the black x-axis error bars.}
    \label{fig:grain-properties}
\end{figure}

In Figure \ref{fig:grain-properties}, we plot the STIS to MIRI F1550C flux ratio for pure astrosilicates, pure organics, and a combination of the two materials including our best-fit composition as a function of scattering angle and porosity. We find that for a range of scattering angles and porosities that a Cat's Tail composed purely of astrosilicates would have a STIS to MIRI F1550C flux ratio of between two to four orders of magnitude higher than for materials made largely of organic refractory material. For example, an astrosilicate-dominated Cat's Tail is 0.1--10$\times$ brighter at STIS wavelengths compared to MIRI F1550C, compared to the measured flux ratio.

For dust grains with a porosity of 0.81 and a composition of $96\%$ organic refractory material and $4\%$ astrosilicates (median posterior values), the Cat's Tail feature is within the STIS detection limits of $0.1\%-0.4\%$ the MIRI F1550C flux at scattering angles $80^\circ-135^\circ$. For scattering angles $<80^\circ$, the STIS to MIRI F1550C flux ratio increases by roughly three orders-of-magnitude, converging at a STIS to MIRI F1550C flux ratio of $\sim1$ as the scattering angles approaches $0^\circ$.

\begin{table*}
\centering
\begin{tabular}{lcl}
\hline\hline
\textbf{Parameter} & \textbf{Description} & \textbf{Value} \\
\hline
\multicolumn{3}{l}{\textbf{Posterior Parameters}} \\
\hline
$f_{sil}$ & Fraction of solids in astronomical silicates             & $0.040_{-0.032}^{+0.106}$ \\ 
  $f_{org}$   & Fraction of solids in organic refractory material      &  $0.960_{-0.106}^{+0.032}$ \\ 
   $P$      &  Porosity of the dust grains       & $0.895_{-0.153}^{+0.067}$  \\
$\theta$ [$^{\circ}$] &  Scattering angle of dust grains in the Cat's Tail       &  $107.63_{-30.37}^{+28.42}$ \\
\hline
\multicolumn{3}{l}{\textbf{Derived Parameters}} \\
\hline
 $\frac{\mathcal{F}_{\mathrm{STIS}}}{\mathcal{F}_{\mathrm{F1550C}}}$     &  Cat's Tail flux ratio at STIS and MIRI F1550C wavelengths & $0.0016_{-0.0008}^{+0.0015}$ \\
   $<T_{dust}>$ [K]&  Size-averaged dust temperature & $139.42_{-18.37}^{+16.49}$       \\
$r(\theta)$ [au]&  True separation of the Cat's Tail midpoint & $183.62_{-15.61}^{+63.18}$    \\

$\frac{\mathcal{F}_{\mathrm{F1550C}}}{\mathcal{F}_{\mathrm{F2300C}}}$     &  Cat's Tail flux ratio at MIRI F1550C and F2300C wavelengths  & $0.470_{-0.177}^{+0.185}$  \\
  \\

\hline
\end{tabular}
\caption{Posterior and derived dust grain properties based on the STIS detection limits of the Cat's Tail flux in scattered light.}
\label{tab:disk_properties}

\end{table*}

\begin{figure}
    \centering
    \includegraphics[width=\linewidth]{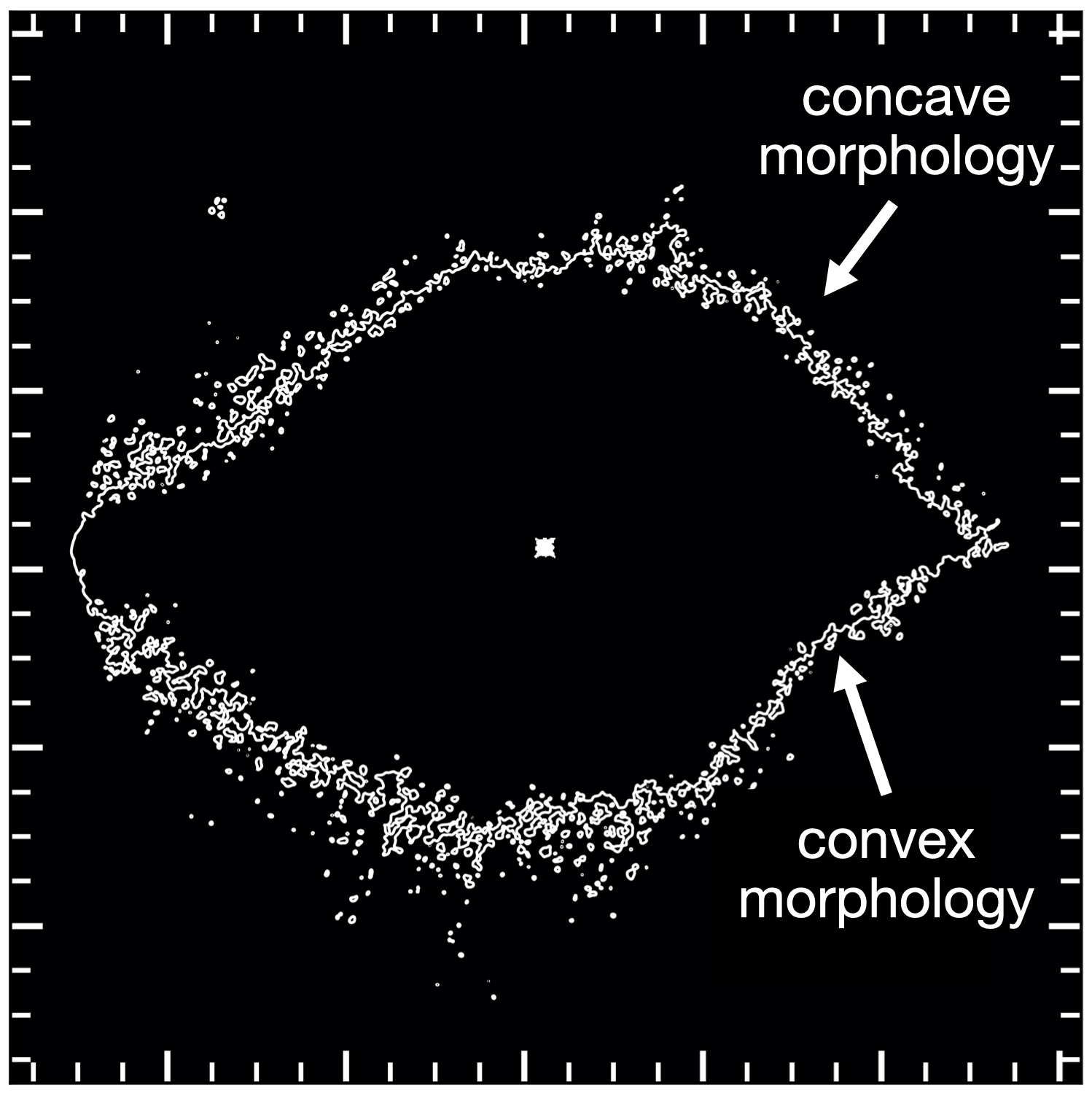}
    \caption{Contours highlighting the morphology of the outer halo of $\beta$ Pic seen with HST/STIS. The arrows highlight the concave/convex morphology described in Section \ref{convave-convex-section}.}
    \label{warp_data_model}
\end{figure}

\section{Discussion}\label{discussion}

\subsection{Apparent Truncation of the Scattered Light Component of the Cat's Tail}\label{sec:truncation}

One important difference between the scattered light and thermal emission morphologies of the Cat's Tail is the apparent truncation of the scattered light component. However, apparent truncation of the Cat's Tail with HST/STIS is not surprising. Since HST/STIS probes light scattered from dust grains, as opposed to thermally emission with MIRI, the surface brightness from light scattered by dust grains is inversely proportional to the square of the distance from the central star. The total separation of the Cat's Tail structure increases along the tail, with the tip hosting grains that are the furthest from the central star, with \cite{Rebollido2024} modeling the tip of the Cat's Tail in MIRI being at separations $>1000$ au. Due to this large separation of at the tip of the tail and the organic refractory material having a low scattering efficiency, it is likely that the apparent truncation is due to the low surface brightness of dust grains at large separations rather than a morphological difference.

\subsection{Utility and Limitations of Scattered Light and Thermal Emission Observations to Constrain the Location of the Cat's Tail}

In our modeling of the Cat's Tail properties, we show how a combination of scattered light and thermal emission observations of the Cat's Tail is able to constrain the location of the substructure relative to the observer. Since the scattering phase function acts to modulate the brightness seen in scattered light, while the mid-IR thermal emission is independent of scattering angle, the flux ratio between the visible/near-IR and mid-IR is able place constraints on the phase function of the dust in the substructure. In the case of the Cat's Tail, we find the location of the Cat's Tail inferred from the flux ratio of visible/near-IR to mid-IR suggests that the Cat's Tail is more likely to be behind the star relative to the observer. However from our analysis and modeling, we are unable to rule out the Cat's Tail being in or near the plane of the sky. 

We note that the Henyey-Greenstein scattering phase function, which is used in this work, is a numerical model that approximates the scattering properties of dust grains and is a crucial first step in understanding the scattering geometry of substructures like the Cat's Tail. Future work incorporating methods like Mie scattering or discrete dipole approximation may yield more conclusive results on the geometry and location of the Cat's Tail substructure, especially if the Cat's Tail is found in multiple new or archival NIRCam filters. However, this approach is beyond the scope of the current work.

\subsection{Sources of the Organic Refractory Material in the Cat's Tail}

The findings of this work and that of \cite{Rebollido2024} point to the Cat's Tail being largely composed of porous organic refractory material. In the Solar System, organic refractory material best reproduces the red surface spectra of Kuiper Belt objects \citep[][]{Cruikshank1998}, most notably Pluto. However, surface tholins make up a negligible amount of the total mass of Pluto, which is not enough to create the $10^{19}$ kg ($\sim$ 0.1$\%$ the mass of Pluto) of small organic refractory dust grains needed to reproduce the Cat's Tail structure in $\beta$ Pic \citep[][]{Rebollido2024}. This points to organic material from the interior of the disrupted planetesimals as the likely origin of the organic refractory material seen in the Cat's Tail. This invites the question: do dwarf planets in exo-Kuiper Belts (like that of $\beta$ Pic) contain enough organic material to create the Cat's Tail? To begin to answer this question, we turn to the organics fractions of comets and icy dwarf planets in our Solar System.

\subsubsection{Abundance of Carbon in the ISM}

In this section, we seek to place constraints on the abundance of refractory organics through observations of the local interstellar medium (ISM). Measurements of the abundance of carbon in the local ISM offer an opportunity to place constraints on the abundance of refractory carbon in extrasolar planetesimals. Studies of interstellar extinction point to $\sim50\%$ of the carbon inventory of the ISM being in gas, while the rest is in refractory dust grains \citep[][]{Zuo2021,Bergin2026}. This yields a C/Si $\sim$ 5 for interstellar grains \citep{Asplund2021,Amarsi2021}. From this constraint we can deduce what fraction of solids in planetesimals are made up of refractory carbon from first principles.

Assuming an atomic mass of 12 amu per carbon atom and a C/Si $\sim$ 5, there are $\sim60$ amu of carbon per silicon atom. Assuming the silicon atoms are all in the form of the most abundant form of pyroxene,  enstatite (MgSiO$_3$), this yields an atomic mass of $\sim$ 100 amu per silicon atom. Dividing the 60 amu per carbon atom by the total mass of refractory carbon and silicates (160 amu), this yields refractory carbon making up $\sim38$ wt$\%$ of interstellar dust. Repeating this exercise for the most common form of olivine, forsterite (Mg$_2$SiO$_4$, $\sim140$ amu), this provides a refractory carbon making up $\sim30$ wt$\%$ of interstellar dust. This gives refractory carbon a range of between $\sim30$ wt$\%$ to $\sim38$ wt$\%$ for grains in the ISM that will potentially coalesce to produce planetesimals. Since we are not taking into account the role of ice in the makeup of outer planetary system planetesimals, we treat this range as an upper limit of the mass fraction of refractory carbon in ISM grains.

\subsubsection{Refractory Organics in the Outer Solar System}

The most direct measurement of the bulk organics fraction in icy bodies came from in-situ measurements of the $\sim$4 km diameter Jupiter-family comet 67P/Churyumov–Gerasimenko (67P) from the \textit{Rosetta} mission. Mass spectrometry of dust grains from 67P found that organics made of 50$\%$ of the of the dust particles by mass \citep[][]{Bardyn2017}. Assuming that the grains sampled by Rosetta are representative of the bulk organics fraction of 67P and assuming a dust-to-ice ratio of 1:1 \citep[][]{Choukroun2020}, it can be deduced that 67P contains $\sim25\%$ organics by mass. However, the progenitors responsible for the Cat's Tail in $\beta$ Pic were estimated to be between 100 and 500 km in radius, the size of dwarf planets in our Solar System \citep[][]{Rebollido2024}. The fraction of organics in dwarf planets is also thought to be substantial, but the bulk organic fraction of dwarf planets is less constrained than for primitive bodies like 67P, since differentiation and varying degrees of thermal and aqueous processing in the interior cause the surface composition to no longer be representative of the bulk composition \citep[e.g.,][]{McKinnon2021}. However, moment of inertia analyses of the dwarf planet Haumea enabled bulk composition estimates, finding that the moment of inertia of the body is best explained by containing 14 wt$\%$ chondritic refractory organic matter \citep[][]{Reynard2023}.

\subsubsection{Size Determination of the Cat's Tail Progenitors Planetesimals}

Combining a Solar System and ISM inspired range of refractory organics in planetesimals, we get a range of 14-38 wt$\%$ for refractory organics in outer planetary systems. From this, we can estimate what the mass of the colliding progenitors must have been to reproduce enough organic refractory material to create the Cat's Tail. Assuming a catastrophic collision between two equal mass progenitors where the smallest fragment has a radius of 0.1 $\mu$m and the largest fragment has a radius of 10 km, we apply a mass-modified form of the Dohnanyi size-distribution ($dM/ds \propto s^{3-\alpha}$), and a steeper-than-Dohnanyi power-law index ($\alpha=3.8$) for a collisional cascade not yet in steady state \citep[][]{Gaspar2012,Leinhardt&Stewart2012}. Using an organic mass fraction of 14-38$\%$ for the parent bodies, we find that each parent body must have had a mass of $\sim1-3\times10^{21}$ kg to produce $2\times 10^{19}$ kg of organic refractory debris in the Cat's Tail, which is in agreement with the upper limit for the progenitor sizes estimated by \cite{Rebollido2024}. Assuming a bulk density of 1.5 g cm$^{-3}$, which is typical of dwarf planets in our Solar System \citep[e.g.,][]{Bierson2019}, this corresponds to a radius of 600-800 km. The mass and radius range is similar to that of dwarf planets in the Kuiper Belt, like Charon and Makemake \citep[][]{Ortiz2012,McKinnon2021}.

We note that a major assumption in the method outlined in this section is that the debris created from the collision has a constant density across all size ranges. We know from \cite{Rebollido2024} and this work that the grains in the Cat's Tail are likely highly porous and must have a density that is significantly smaller than the larger debris fragments created from the collision due to mechanisms like gravitational compacting \citep[e.g.,][]{Bierson2019}. For this reason, the estimated mass of $1-3\times10^{21}$ kg should be treated as a lower limit for the mass of each colliding progenitor. 

\subsection{Where are the Silicates from the Collision that Created the Cat's Tail?}

In addition to an inventory of organic refractory material, icy moons and dwarf planets in our Solar System are thought to also host a large inventory of silicates in their interior \citep[e.g.,][]{Olkin2017,Reynard2023}. For example, \cite{Reynard2023} determines a silicate mass fraction of 45-70$\%$ for all icy moons and dwarf planets where moment of inertia analyses are possible, compared to our estimate of 14-38$\%$ for the organic refractory material. However, the Cat's Tail structure of $\beta$ Pic, thought to be created from the collisional aftermath of dwarf planet mass progenitors, contains almost entirely organic refractory material and little to no silicates. In this section we will explore where organics-rich collisional remnants are possible for the Cat's Tail and where the silicate debris from this collision could be. 

While the location of organics within the interior of icy moons and dwarf planets in our Solar System is relatively unconstrained, recent studies have inferred from moment of inertia analyses that a majority of the organics inventory of the bodies is mixed into the silicate cores \citep[e.g.,][]{Olkin2017,Neri2020}. Given the large amount of organics located in the silicate core of planetesimals and the estimate of $10^{19}$ kg of organic refractory material thought to be in the Cat's Tail \citep{Rebollido2024}, it is likely that the break-up of the planetesimal responsible for the Cat's Tail was catastrophic in order to produce the amount of organics observed. However, without invoking a carbon dwarf planet, this implies that the silicates from the core should have also been released after the collision alongside the organic refractory material, but silicate-dominated substructures are not distinctly identified in the MIRI imaging.

\subsubsection{Radiation Pressure-Driven Spatial Differentiation of Silicates and Organic Refractory Material}

Radiation pressure-driven differentiation of the collisional debris is one possible mechanism that explains the organics-dominated Cat's Tail. A key difference of astrosilicates and organic refractory material is their associated $\beta$ value that governs the dynamics of the grain once created, with the astrosilicates having smaller $\beta$ values than organics for a given size $s$ \citep[e.g.,][]{Draine1984,Li&Greenberg1997}. For this reason, it could be that the silicates created from the aftermath of the collision that created the Cat's Tail occupy different spatial regimes within $\beta$ Pic compared to the organics created from the collision. Specifically, \cite{Rebollido2024} finds that highly porous silicates have a 0.5 < $\beta$ < 1 putting them in the hyperbolic blowout regime, while the porous organics have $\beta$ > 1 out to grain sizes of > 20 $\mu$m putting the grains in the anomalous hyperbolic blowout regime. For both silicate and organic blowout grains, this difference would result in different spatial distributions of each grain type, with redder regions of the nebulosity seen with MIRI potentially being silicate dominated, while the bluer regions being organics dominated (see color map of $\beta$ Pic with MIRI in Figure 8 of \citealt{Rebollido2024}). Furthermore, the larger and/or highly porous silicates created from the collision event may remain loosely bound to the system, being located in bound substructures like the SW dust clump and the extended secondary disk. \cite{Telesco2005} find that the SED of the SW dust clump is distinct from that of the rest of the disk, implying a size and/or compositional difference of dust in the clump. While we are not suggesting that the out-of-midplane substructures seen in the mid-IR are purely composed of silicates, they could be composed of larger quantities of silicates in addition to organics than the Cat's Tail.

\subsection{Origins of $\beta$ Pic's Asymmetric/Warped Dust Halo}\label{warp-discussion}


The origin of the warp observed in the midplane has been classically attributed to secular forcing of planetesimals and, in turn, dust by the inclined orbit of $\beta$ Pic b \citep[e.g.][]{Mouillet1997,Augereau2001}. That said, it remains unclear how the presence of the two additional planets, $\beta$ Pic c and d, discovered after the original theoretical studies, affects this interpretation. Nevertheless, there is some evidence that $\beta$ Pic b remains the dominant perturber \cite[e.g.][]{Dawson2011,Nesvold&Kuchner2015,Matra2019,Smallwood2023}. The observed convex/concave morphology of the outer halo of $\beta$ Pic is reminiscent of the warped morphology observed in the midplane of $\beta$ Pic \citep[e.g.][]{Golimowski2006,Apai2015}, where the SW side of the midplane is vertically extended above/north of the rest of the midplane, and the NE midplane vertically extended below/south of the rest of the midplane.

The shared morphology of the asymmetries in the midplane and halo suggests that $\beta$ Pic b may be responsible for both. Theoretical investigations of an inclined planet interacting secularly with an exterior debris disk predict distinct vertical structures depending on the disk-to-planet mass ratio $M_d/m_p$  \citep[][]{Sefilian2025}. In particular, for a massless disk, the planet produces a warp broadly consistent with that observed in the midplane, but the resulting box-like disk morphology does not reproduce the detailed structure of the halo (\cite{Nesvold&Kuchner2015}; see Figure 6 in \citet{Sefilian2025}). At the other extreme, when $M_d/m_p \gtrsim 1 $, the disk's gravity suppresses the warp, leaving the entire disk nearly razor-thin and misaligned with the planetary orbit; a regime that can be ruled out for $\beta$ Pic. The observed convex/concave halo morphology instead appears to be more closely reproduced in the intermediate regime, namely, when the disk has a small but non-zero mass such that $M_d/m_p \lesssim 1$ (see Figure 9 in \citet{Sefilian2025}). Interestingly, in this regime the disk also develops a curved, X-like structure that is reminiscent of the Cat's Tail. Considering $\beta$ Pic b only, such an outcome is expected for the $\beta$ Pic disk provided its total mass remains below $\sim 10^3$ $M_{\earth}$ (see Figure 12 in \citet{Sefilian2025}). 


While these similarities are suggestive, several important caveats remain when comparing the observed morphology with the structures predicted in \cite{Sefilian2025}. For example, the curvy X-like structure appears on both sides of the star in their model, unlike the one-sided Cat's Tail observed in $\beta$ Pic. Moreover, the theoretical investigation in \cite{Sefilian2025} concerns the gravitational evolution of large planetesimals that are unaffected by radiation pressure, whereas the small grains traced by HST/STIS lie in the photogravitational regime. It therefore remains unclear how directly the predicted structures map onto the small-grain halo observed here, or why similar features are not apparent at longer wavelengths, such as with ALMA. Accordingly, a definitive interpretation will require theoretical modeling that accounts for the full planetary system and radiation pressure. Such modeling is beyond the scope of this work, but the qualitative similarities nevertheless suggest that disk gravity may play an important role in shaping the $\beta$ Pic debris disk. 

\subsection{Comparison of Asymmetric Halo Observed with HST/STIS and Previously Observed Asymmetries}

The debris disk around $\beta$ Pic exhibits several well-studied asymmetries both at small and large separations from the central star. Among the most prominent is the butterfly asymmetry, an out-of-midplane surface-brightness asymmetry observed at large separations \citep[e.g.,][]{Kalas&Jewitt1995}. We recover this feature in the halo of $\beta$ Pic in the HST/STIS observations (see left panel of Figure \ref{fig:halo_asymmetries}). This butterfly asymmetry has been attributed to an extension of the midplane warp, where the dust produced in the warped midplane is subsequently blown out by radiation pressure, producing the surface brightness asymmetry in the halo \citep[e.g.,][]{Golimowski2006,Apai2015}.

However, radiation pressure alone cannot account for the warped, convex/concave morphology of the halo seen in our HST/STIS observations (see Figure \ref{warp_data_model}). The secular perturbation of planetesimals and dust is required to recreate the convex/concave or warped morphology in the halo, as discussed in Section \ref{warp-discussion}. However, this still suggests that both features are the result of the inclined inner planet $\beta$ Pic b (or now d). The butterfly surface brightness asymmetry requires a warped midplane to feed excess dust into certain lobes of the halo, and the secular perturbation of planetesimals/dust by an inclined planet is required to recreate to observed vertical structure of the halo.

While large field-of-view observations from the ground in \cite{Kalas&Jewitt1995} show a slight convex extension of the halo of $\beta$ Pic, the complete warped morphology seen with HST/STIS is not as apparent. We attribute this to the difference in angular resolution of HST/STIS compared to the angular resolution of the ground-based observations. The seeing-limited PSF of the ground-based observations (FWHM $\sim$ 1.4 arcsec, $\sim$27 au for $\beta$ Pic) likely convolves the surrounding halo on scales comparable to the warped structure itself, while the HST/STIS PSF (FWHM $\sim$ 0.07 arcsec, $\sim$1.4 au for $\beta$ Pic) is able to resolve the warped morphology.

\section{Summary}

In this study, we present a combination of archival and new epochs of HST/STIS coronagraphic observations of the $\beta$ Pictoris debris disk. The key findings are as follows:

\bigskip

1) The debris disk is detected out to a projected separation of $\sim$500 au for the first time in HST/STIS coronagraphic imaging. We median combine HST/STIS images of $\beta$ Pic spanning 14 years and a maximum pixel summed exposure time of $> 10000$ seconds, providing the deepest image of $\beta$ Pic with HST/STIS. We detect the disk at high SNR throughout the midplane, with the SNR at separations of 50-200 au being consistently greater than 200. 

2) We search for and detect (SNR$\sim$5--10) the visible/near-infrared scattered light component of the "Cat's Tail" collisional remnant discovered with MIRI imaging.

3) We detect a structure that extends past the outer halo of the disk that is in the same region of the extended secondary disk discovered with MIRI. However, the low SNR nature and slight offset from that of the extended secondary disk in MIRI makes it difficult to rule out the observed structure being a PSF artifact.

4) Using the MIRI F1550C image of $\beta$ Pic from \cite{Rebollido2024}, we create thin models of the Cat's Tail that is injectable into the presented HST/STIS image. We find that the flux of the Cat's Tail at HST/STIS wavelengths must be between $0.1\%$ and $0.4\%$ of its flux at F1550C wavelengths, assuming the same morphology. 

5) Using the STIS flux limit of $0.1-0.4\%$ the F1550C Cat's Tail flux in addition to prior constraints set by \cite{Rebollido2024}, we model the properties of the dust grains inside the Cat's Tail structure. We find that that the fraction of silicates only make up $<14\%$ of the composition of the grains in the Cat's Tail, with the rest being composed of organic refractory material. We find that the grains must be highly porous, with a porosity of 0.55 -- 0.93, to explain the detection with HST/STIS, which is consistent with the findings of \cite{Rebollido2024}. Finally, we find that scattering angles of the dust in the Cat's Tail must be $80^\circ-135^\circ$ to match the flux range observed with HST/STIS. 

6) Using the mass of organics in the Cat's Tail calculated by \cite{Rebollido2024}, we estimate the mass of each colliding progenitor, informed by the organics mass fraction of Solar System dwarf planets and planetesimals, that is required to produce the Cat's Tail. We set a lower limit mass of at least $1-3\times 10^{21}$ kg per colliding progenitor, comparable to Charon and Makemake in our Solar System.

7) We also find a convex/concave morphological asymmetry in the outer halo of $\beta$ Pic with HST/STIS. We argue that the morphological asymmetry closely resembles the vertical structure produced by secular interactions between an inclined planet and a less massive exterior disk \citep{Sefilian2025}. However, dedicated follow-up dynamical modeling is needed to confirm this.

\section*{Acknowledgments}

The results reported herein benefited from collaborations and/or information exchange within NASA’s Nexus for Exoplanet System Science (NExSS) research coordination network sponsored by NASA’s Science Mission Directorate. This material is based on work supported by grants HST-GO-16174, HST-GO-17456, and HST-GO-17741 obtained at the Space Telescope Science Institute (STScI), which is operated by the Association of Universities for Research in Astronomy, Inc., under NASA contract NAS 5-26555, and the National Aeronautics and Space Administration under agreement No. 80NSSC21K0593 for the program “Alien Earths.” A.A.S. is supported by the Heising-Simons Foundation through a 51 Pegasi b Fellowship.

The HST/STIS data presented in this work were obtained from the Mikulski Archive for Space Telescopes (MAST) at the Space Telescope Science Institute. The specific observations used in this article can be accessed via \dataset[doi: 10.17909/hr6e-dv84]{https://doi.org/10.17909/hr6e-dv84}.

\appendix

\section{Details Regarding the HST/STIS Coronagraphic Observations of $\beta$ Pic}\label{sec:STIS_obs}

In this section, we highlight the observing configuration across all epochs of data used in the final image of $\beta$ Pic in this work. The details regarding the observations of the $\beta$ Pic inner disk can be found in Table \ref{tab:obs}, and the details regarding the observations of the $\beta$ Pic outer disk can be found in Table \ref{tab:obs-2}. We note that visits 55-59 in Table \ref{tab:obs-2} are HOPR repeat observations for guide star failures in the previous set of observations. 

\begin{figure*}
    \centering
    \includegraphics[width=\linewidth]{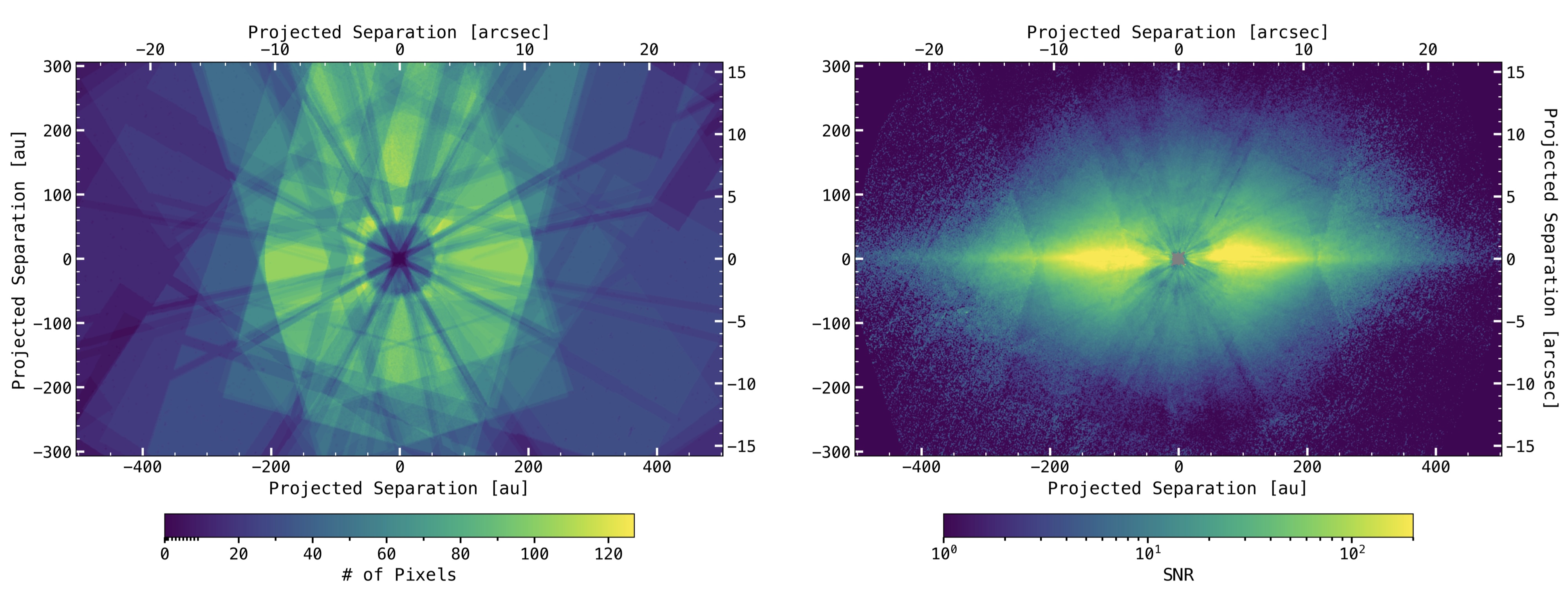}
    \caption{\textit{Left:} Map depicting the number of images that were combined in each pixel of the final combined image. \textit{Right:} SNR map of the combined $\beta$ Pic image presented in this work. We detect the disk at very high SNR, especially between 50 to 150 au in projected separation.}
    \label{fig:npix-SNR}
\end{figure*}

The SNR and number of observations per pixel map can be found in Figure \ref{fig:npix-SNR}. We note our highest SNR region is between 50 and 200 au. That is the approximate location of the parent-body belt, in addition to there being more observations of this region between all of the epochs median combined to create our final image. This is apparent in the map of the number of observations per pixel.

\section{Verification of Scattered Light Cat's Tail Feature}\label{sec:jackknife}

\begin{figure*}
    \centering
    \includegraphics[width=\linewidth]{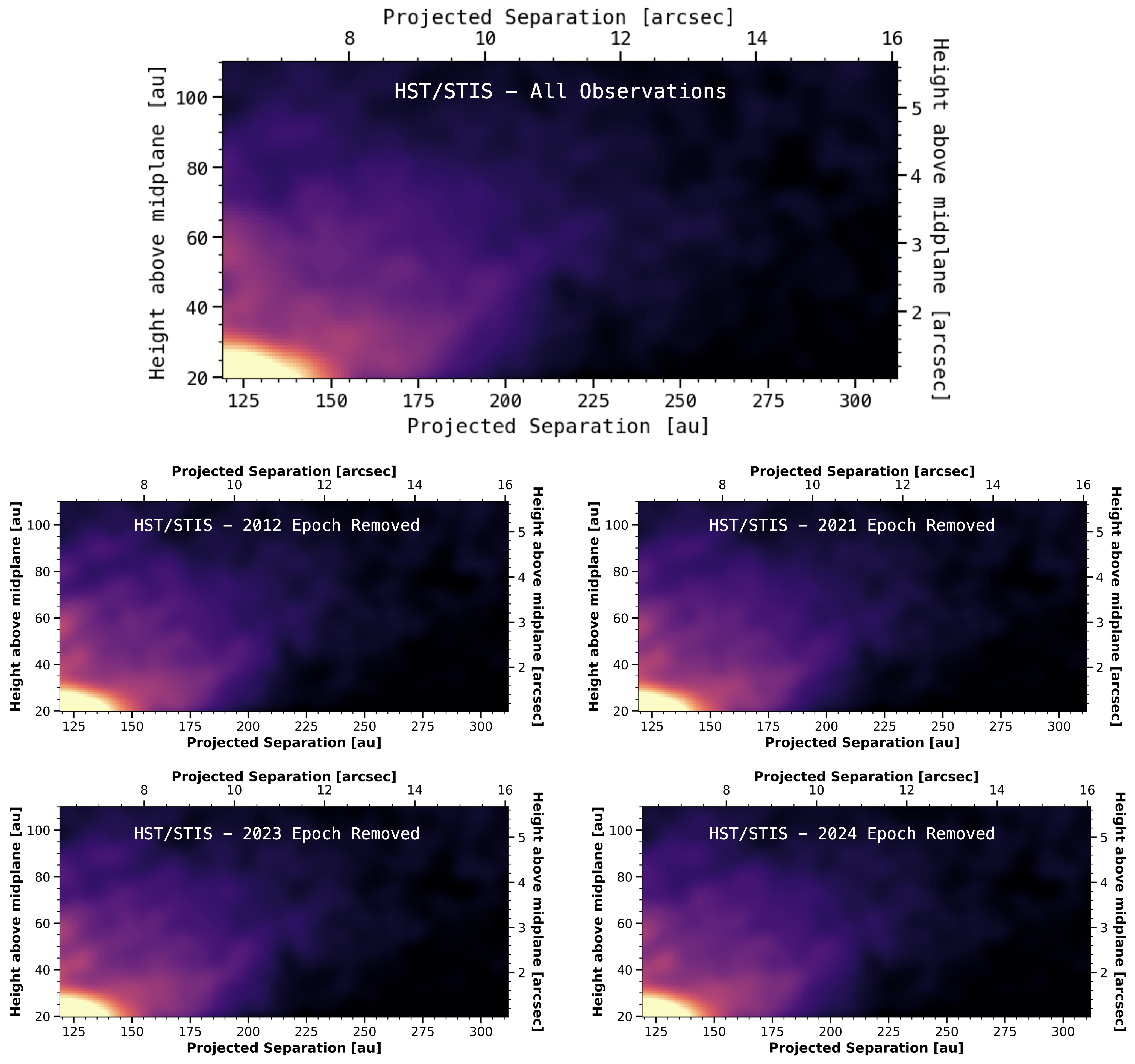}
    \caption{Jackknife test of the Cat's Tail feature in the HST/STIS observations. The top image shows the the combined Cat's Tail detection with all observations used. The panels below show the Cat's Tail with one epoch removed from the combined image, highlighting that the Cat's Tail feature we see in the HST/STIS observations are unlikely to be an artifact introduced by a single epoch of data. }
    \label{fig:jackknife}
\end{figure*}

In this section, we seek to confirm that the detection of the Cat's Tail is not an artifact of a single epoch of HST/STIS data. To do this, we conduct a jackknife test, where we remove a single epoch of data (see Section \ref{cats-tail-section}) and confirm that the Cat's Tail feature does not disappear. The results of the jackknife test can be found in Figure \ref{fig:jackknife}. We find that, although the feature appears noisier due to the reduction of observations in the region, especially at the tip, we recover the Cat's Tail in each set of images where an epoch of data was removed. While the detection of the Cat's Tail is relatively low S/N ($\sim$5-10) in our data, this test shows that the feature we attribute to the scattered light component of the Cat's Tail is unlikely to be an artifact introduced by the systematics of a single epoch of observations.

\section{Astrometry of S1}\label{sec:S1-astrometry}

In this section, we will outline the astrometry conducted on the S1 point source. Due to S1 being at a projected separation of 400 au, it is only probed in the observations of the outer disk taken as part of GO-17456 (PI: Wagner). The observations were taken from late 2023 to late 2025, with three distinct periods of observations where the region of the disk at separations where S1 is visible are probed (2023.87, 2024.78, 2025.81). 

In order to distinguish between S1 being a background source or an object gravitationally bound to $\beta$ Pic, we will compare the astrometry of S1 to the parallactic and proper motion of $\beta$ Pic. If S1 is a background star, it will move in the opposite direction of $\beta$ Pic's proper motion.

$\beta$ Pic is a high proper motion target with a proper motion of 4.65 $\pm$ 0.11 mas yr$^{-1}$ in right ascension and 83.10 $\pm$ 0.15 mas yr$^{-1}$ in declination \citep[][]{Snellen2018}. Assuming the image is oriented in the "north up, east left" direction and a STIS detector plate scale of 50.7 mas pixel$^{-1}$ \citep[][]{STIS-DG}, this translates to 0.092 $\pm$ 0.002 pixels yr$^{-1}$ in the $x$ direction and --1.639 $\pm$ 0.003 pixels yr$^{-1}$ in the $y$ direction in the STIS image of $\beta$ Pic. The parallax of $\beta$ Pic is 51.44 $\pm$ 0.12 mas \citep[][]{Snellen2018}. The expected relative motion of S1 as a background star can be found in Figure \ref{fig:astrometry}.

\subsection{Fitting the Relative Position of S1 from 2023 to 2025}

\begin{figure*}
    \centering
    \includegraphics[width=\linewidth]{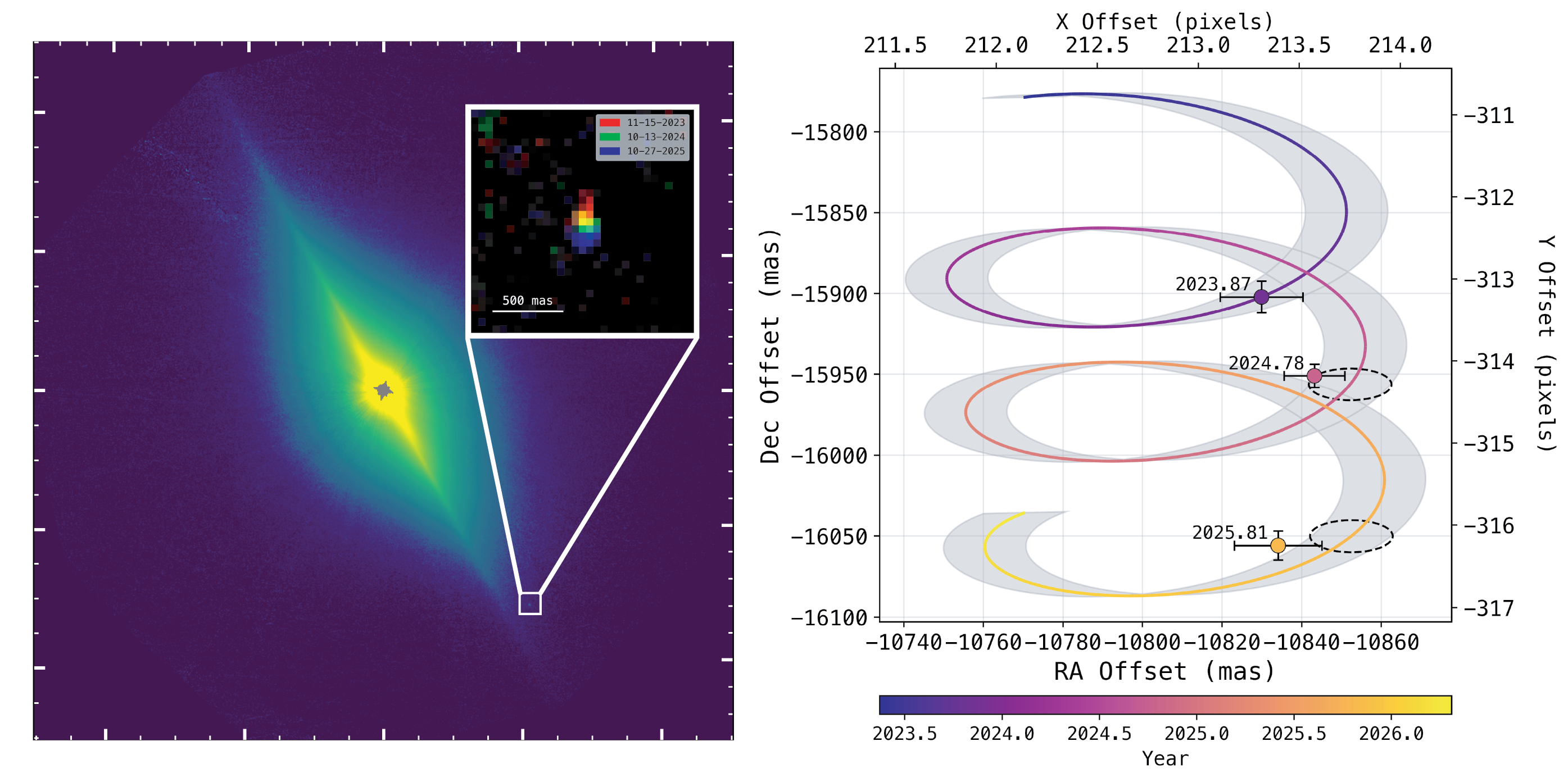}
    \caption{\textit{Left:} RGB cutout of the three epochs of data where S1 was within the field-of-view, showing the change in position of S1 throughout the epochs. \textit{Right:} Astrometric measurements of S1 from 2023 to 2025. The solid line indicates the expected motion of a background source over time. The dashed black circles indicate the expected location of S1 at the times the observations were taken if it is a background source. As seen, the position of S1 in subsequent observations appears in the expected position of a background source within uncertainty. }
    \label{fig:astrometry}
\end{figure*}

To find the center and associated uncertainties of S1 in each epoch (2023.87, 2024.78, 2025.81), we inject negatives of simulated STIS PSFs and vary $x$, $y$, and intensity PSF parameters. The resulting astrometric measurements of S1 in each epoch are shown in Figure \ref{fig:astrometry}. We find that the astrometry of S1 in the two years of observations we present is consistent with that of a background star. To confirm this result, we attempt to fit an orbit to our astrometric measurements of S1 using {\tt\string orbitize!} \citep[][]{Blunt2020}. Using the OFTI mode, we are unable to fit any Keplerian orbits in the case that S1 may be bound to $\beta$ Pic, reaffirming S1 as a background star.

\section{Modeling Dust Grain Properties with Different Silicate Optical Constants}\label{pyox-olivine}

Our modeling of the grain properties in the Cat's Tail adopts the optical constants of astronomical silicates \citep{Draine1984}. An astronomical silicate is not a measured mineral but a synthetic set of optical constants assembled to reproduce interstellar extinction. To assess whether our retrieved grain properties for the amounts of organics depends on this choice of silicate, we repeat the model fitting of the Cat's Tail grain properties (see Sections \ref{model-fitting-section} and \ref{model-results-section}) using optical constants for silicates like olivine and pyroxene \citep{Dorschner1995}.

\begin{figure*}[ht!]
    \centering
    \includegraphics[width=\linewidth]{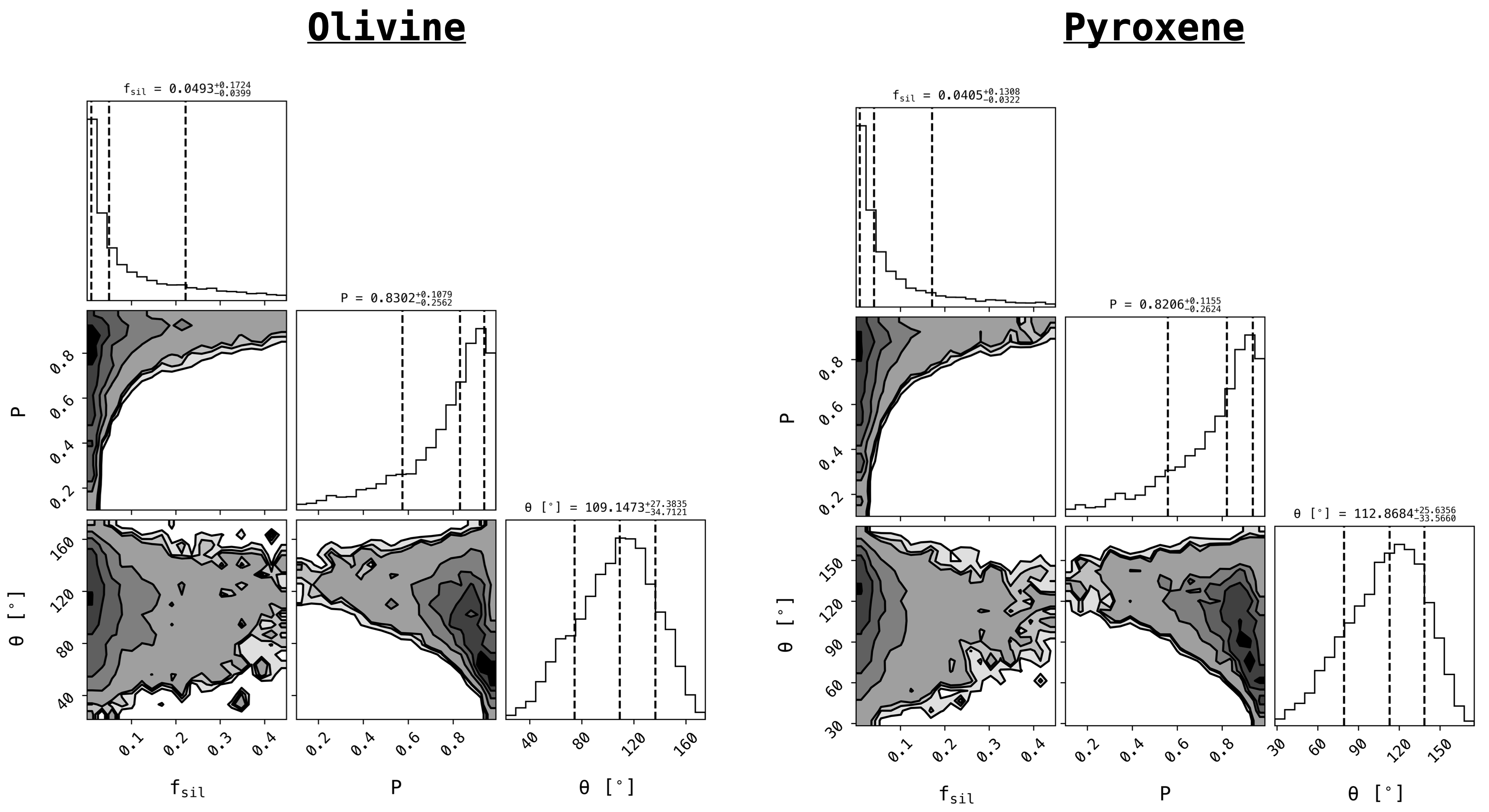}
    \caption{The posterior distribution of the {\tt\string emcee} chain of the three dust grain parameters of the Cat's Tail using the optical constants of olivine and pyroxene \citep[][]{Dorschner1995} instead of astronomical silicates. The model is fit with 16 walkers, 5000 steps, and 500 burn-in steps. The dashed lines indicate the 16th, 50th, and 84th percentiles, respectively.}
    \label{fig:other-mcmc}
\end{figure*}

We find that the choice of underlying silicate does not change the grain properties retrieved. With the three parameters that were fit (silicate fraction, porosity, and scattering angle), we find using astrosilicates is consistent within one-sigma compared to grain properties fit using the optical properties of olivine and pyroxene (see Figure \ref{fig:other-mcmc}). This indicates that these fit grain properties are robust against the choice of silicate optical constants.

\begin{deluxetable*}{cccccc}
\tabletypesize{\footnotesize}
\tablecolumns{6}
\tablewidth{\textwidth}   
\tablecaption{ Summary of the HST/STIS observations of \bpic{} and \apic{} for the inner disk observations from 2012 to 2025.}
\tabletypesize{\scriptsize} 
\label{tab:obs}
\tablehead{
\colhead{Program}  & \colhead{Date} &  \colhead{Visit \#} & \colhead{Target} & \colhead{Int. Time [s]} & \colhead{Occulter}\\}
\startdata
12551 & 03/06/2012 & 1 & \bpic{} &$11\times1.2$ & WEDGEA0.6 \\ 
12551 & 03/06/2012 & 1 & \bpic{} &$4\times60.0$,$16\times3.0$ & WEDGEA1.0 \\ 
12551 & 03/06/2012 & 1 & \bpic{} & $4\times 60.0$, $16\times3.0$ & WEDGEB1.0 \\ 
12551 & 03/06/2012 & 1 & \bpic{} & $11\times1.2$ & WEDGEB0.6 \\
12551 & 03/06/2012 & 2 & \apic{} & $11\times0.7$ & WEDGEA0.6 \\ 
12551 & 03/06/2012 & 2 & \apic{} & $4\times36.0$ ,$16\times1.9$ & WEDGEA1.0 \\
12551 & 03/06/2012 & 2 & \apic{} & $4\times36.0$, $16\times1.9$ & WEDGEB1.0 \\ 
12551 & 03/06/2012 & 2 & \apic{} & $17\times0.7$ & WEDGEB0.6 \\ 
12551 & 03/06/2012 & 3 & \bpic{} &$11\times1.2$ & WEDGEA0.6 \\ 
12551 & 03/06/2012 & 3 & \bpic{} &$4\times60.0$,$16\times3.0$ & WEDGEA1.0 \\ 
12551 & 03/06/2012 & 3 & \bpic{} & $4\times 60.0$, $16\times3.0$ & WEDGEB1.0 \\ 
12551 & 03/06/2012 & 3 & \bpic{} & $11\times1.2$ & WEDGEB0.6 \\
\hline
16788 & 02/28/2021 & 1 & \bpic{} & $8\times1.2$ & WEDGEA0.6 \\
16788 & 02/28/2021 & 1 & \bpic{} & $4\times60.0$, $16\times3.0$ & WEDGEA1.0 \\ 
16788 & 02/28/2021 & 1 & \bpic{} & $4\times60.0$, $16\times3.0$ & WEDGEB1.0 \\ 
16788 & 02/28/2021 & 1 & \bpic{} & $9\times1.2$ & WEDGEB0.6 \\ 
16788 & 02/28/2021 & 2 & \apic{} & $11\times0.7$ & WEDGEA0.6 \\ 
16788 & 02/28/2021 & 2 & \apic{} & $4\times36.0$, $16\times1.9$ & WEDGEA1.0 \\ 
16788 & 02/28/2021 & 2 & \apic{} & $4\times36.0$, $15\times1.9$ & WEDGEB1.0 \\ 
16788 & 02/28/2021 & 2 & \apic{} & $14\times0.7$ & WEDGEB0.6 \\
16788 & 02/28/2021 & 3 & \bpic{} & $9\times1.2$ & WEDGEA0.6 \\ 
16788 & 02/28/2021 & 3 & \bpic{} & $4\times60.0$, $16\times3.0$ & WEDGEA1.0 \\ 
16788 & 02/28/2021 & 3 & \bpic{} & $4\times60.0$, $16\times3.0$ & WEDGEB1.0 \\ 
16788 & 02/28/2021 & 3 & \bpic{} & $8\times1.2$ & WEDGEB0.6 \\ 
\hline
16788 & 03/06/2023 & 1 & \bpic{} & $8\times1.2$ & WEDGEA0.6 \\ 
16788 & 03/06/2023 & 1 & \bpic{} & $4\times60.0$, $16\times3.0$ & WEDGEA1.0 \\ 
16788 & 03/06/2023 & 1 & \bpic{} & $4\times60.0$, $16\times3.0$ & WEDGEB1.0 \\ 
16788 & 03/06/2023 & 1 & \bpic{} & $7\times1.2$ & WEDGEB0.6 \\ 
16788 & 03/06/2023 & 2 & \apic{} & $11\times0.7$ & WEDGEA0.6 \\ 
16788 & 03/06/2023 & 2 & \apic{} & $4\times36.0$, $16\times3.0$ & WEDGEA1.0 \\ 
16788 & 03/06/2023 & 2 & \apic{} & $4\times36.0$, $16\times3.0$ & WEDGEB1.0 \\ 
16788 & 03/06/2023 & 2 & \apic{} & $14\times0.7$ & WEDGEB0.6 \\ 
16788 & 03/06/2023 & 3 & \bpic{} & $9\times1.2$ & WEDGEA0.6 \\ 
16788 & 03/06/2023 & 3 & \bpic{} & $4\times60.0$, $16\times3.0$ & WEDGEA1.0 \\ 
16788 & 03/06/2023 & 3 & \bpic{} & $4\times60.0$, $16\times3.0$ & WEDGEB1.0 \\ 
\hline
16788 & 02/26/2024 & 1 & \bpic{} & $8\times1.2$ & WEDGEA0.6 \\ 
16788 & 02/26/2024 & 1 & \bpic{} & $4\times60.0$, $16\times3.0$ & WEDGEA1.0 \\ 
16788 & 02/26/2024 & 1 & \bpic{} & $4\times60.0$, $16\times3.0$ & WEDGEB1.0 \\ 
16788 & 02/26/2024 & 1 & \bpic{} & $9\times1.2$ & WEDGEB0.6 \\ 
16788 & 02/26/2024 & 2 & \apic{} & $11\times0.7$ & WEDGEA0.6 \\ 
16788 & 02/26/2024 & 2 & \apic{} & $4\times36.0$, $16\times3.0$ & WEDGEA1.0 \\ 
16788 & 02/26/2024& 2 & \apic{} & $4\times36.0$, $16\times3.0$ & WEDGEB1.0 \\ 
16788 & 02/26/2024 & 2 & \apic{} & $14\times0.7$ & WEDGEB0.6 \\ 
16788 & 02/26/2024 & 3 & \bpic{} & $9\times1.2$ & WEDGEA0.6 \\ 
16788 & 02/26/2024 & 3 & \bpic{} & $4\times60.0$, $16\times3.0$ & WEDGEA1.0 \\ 
16788 & 02/26/2024 & 3 & \bpic{} & $4\times60.0$, $16\times3.0$ & WEDGEB1.0 \\ 
16788 & 02/26/2024 & 3 & \bpic{} & $8\times1.2$ & WEDGEB0.6 \\
\hline
17741 & 12/07/2025 & 1 & \bpic{} & $16\times1.2$, $8\times60.0$, $27\times3.0$ & WEDGEA1.0 \\ 
17741 & 12/07/2025 & 2 & \apic{} & $28\times0.7$, $8\times36.0$, $27\times1.9$ & WEDGEA1.0 \\ 
17741 & 12/07/2025 & 3 & \bpic{} & $16\times1.2$, $8\times60.0$, $27\times3.0$ & WEDGEA1.0 \\ 
\enddata
\vspace{-0.8cm}
\end{deluxetable*}

\begin{deluxetable*}{cccccc}
\tabletypesize{\footnotesize}
\tablecolumns{6}
\tablewidth{\textwidth} 
  \tablecaption{Summary of the HST/STIS observations of $\beta$ Pic and
    $\alpha$ Pic for the outer disk observations from 2023 to 2025.
    Visits 55-59 are HOPR repeats of visits 5-9.}
\tabletypesize{\scriptsize} 
\label{tab:obs-2}
\tablehead{
\colhead{Program}  & \colhead{Date} &  \colhead{Visit \#} & \colhead{Target} & \colhead{Int. Time [s]} & \colhead{Occulter}\\}
\startdata
  17456 & 04/30/2025  & 1  & $\beta$ Pic  & 4 $\times$ 60.0 & WEDGEA1.8 \\
  17456 & 04/30/2025  & 2  & $\alpha$ Pic & 4 $\times$ 60.0 & WEDGEA1.8 \\
  17456 & 04/30/2025  & 3  & $\beta$ Pic  & 4 $\times$ 60.0 & WEDGEA1.8 \\
  17456 & 04/03/2024 & 4  & $\beta$ Pic  & 4 $\times$ 60.0 & WEDGEB1.8 \\
  17456 & 04/03/2024  & 5 & $\alpha$ Pic & 4 $\times$ 60.0 & WEDGEB1.8 \\
  17456 & 05/01/2025 & 55 & $\alpha$ Pic & 4 $\times$ 60.0 & WEDGEB1.8 \\
  17456 & 04/03/2024  & 6 & $\beta$ Pic & 4 $\times$ 60.0 & WEDGEB1.8 \\
  17456 & 05/01/2025 & 56 & $\beta$ Pic  & 4 $\times$ 60.0 & WEDGEB1.8 \\
  17456 & 11/15/2023  & 7 & $\beta$ Pic & 4 $\times$ 60.0 & WEDGEA1.8 \\
  17456 & 10/27/2025 & 57 & $\beta$ Pic  & 4 $\times$ 60.0 & WEDGEA1.8 \\
  17456 & 11/15/2023  & 8 & $\alpha$ Pic & 4 $\times$ 60.0 & WEDGEA1.8 \\
  17456 & 10/27/2025 & 58 & $\alpha$ Pic & 4 $\times$ 60.0 & WEDGEA1.8 \\
  17456 & 11/15/2023  & 9 & $\beta$ Pic & 4 $\times$ 60.0 & WEDGEA1.8 \\
  17456 & 10/27/2025 & 59 & $\beta$ Pic  & 4 $\times$ 60.0 & WEDGEA1.8 \\
  17456 & 10/13/2024 & 10 & $\beta$ Pic  & 4 $\times$ 60.0 & WEDGEB1.8 \\
  17456 & 10/13/2024 & 11 & $\alpha$ Pic & 4 $\times$ 60.0 & WEDGEB1.8 \\
  17456 & 10/13/2024 & 12 & $\beta$ Pic  & 4 $\times$ 60.0 & WEDGEB1.8 \\
  \enddata
\end{deluxetable*}

\bibliography{sample7}
\bibliographystyle{aasjournal}

\end{document}